\documentclass[english,preview]{aastex}
\usepackage{verbatim}
\usepackage{textcomp}
\usepackage{amstext}
\usepackage{graphicx}
\usepackage{amsmath}

\makeatletter

\shorttitle{A Parallel MC Code for Collisional $N$-body
Systems}
\shortauthors{Pattabiraman}

\makeatother

\usepackage{babel}
\begin{document}

\title{A Parallel Monte Carlo Code for Simulating Collisional $N$-body
Systems}

\author{Bharath Pattabiraman\altaffilmark{1,2}}

\email{bharath@u.northwestern.edu}

\author{Stefan Umbreit\altaffilmark{1,3}}

\author{Wei-keng Liao\altaffilmark{1,2}}

\author{Alok Choudhary\altaffilmark{1,2}}

\author{Vassiliki Kalogera\altaffilmark{1,3}}

\author{Gokhan Memik\altaffilmark{1,2}}

\author{}

\author{}

\and{}

\author{Frederic A. Rasio\altaffilmark{1,3}}

\altaffiltext{1}{Center for Interdisciplinary Exploration and Research in Astrophysics,
Northwestern University, Evanston, USA.}

\altaffiltext{2}{Department of Electrical Engineering and Computer Science, Northwestern
University, Evanston, USA.}

\altaffiltext{3}{Department of Physics and Astronomy, Northwestern University,
Evanston, USA. }
\begin{abstract}
We present a new parallel code for computing the dynamical evolution
of collisional $N$-body systems with up to $N\sim10^{7}$ particles.  Our code
is based on the H\'enon Monte Carlo method for solving the Fokker-Planck
equation, and makes assumptions of spherical symmetry and dynamical
equilibrium. The principal algorithmic developments involve optimizing data
structures, and the introduction of a parallel random number generation scheme,
as well as a parallel sorting algorithm, required to find nearest neighbors for
interactions and to compute the gravitational potential. The new algorithms  we
introduce along with our choice of decomposition scheme minimize communication
costs and ensure optimal distribution of data and workload among the processing
units. Our implementation uses the Message Passing Interface (MPI) library for
communication, which makes it portable to many different supercomputing
architectures. We validate the code by calculating the evolution of clusters
with initial Plummer distribution functions up to core collapse with the number
of stars, $N$, spanning three orders of magnitude, from $10^{5}$ to $10^{7}$.
We find that our results are in good agreement with self-similar core-collapse
solutions, and the core collapse times  generally agree with expectations from
the literature. Also, we observe good total energy conservation, within
$\lesssim 0.04\%$ throughout all simulations. We analyze the performance of the
code, and demonstrate  near-linear scaling of the runtime with the number of
processors up to 64 processors for $N=10^{5}$, 128 for $N=10^{6}$ and 256 for
$N=10^{7}$.  The runtime reaches saturation with the addition of processors
beyond these limits, which is a characteristic of the parallel sorting
algorithm. The resulting maximum speedups we achieve are approximately
60\texttimes{}, 100\texttimes{}, and 220\texttimes{}, respectively.
\end{abstract}

\keywords{Methods: numerical, Galaxies: star clusters: general, globular
clusters: general, Stars: kinematics and dynamics}

\section{Introduction}

The dynamical evolution of dense star clusters is a problem of fundamental
importance in theoretical astrophysics. Important examples of star
clusters include globular clusters, spherical systems containing typically
$10^{5}$ - $10^{7}$ stars within radii of just a few parsec, and
galactic nuclei, even denser systems with up to $10^{9}$ stars contained
in similarly small volumes, and often surrounding a supermassive black
hole at the center. Studying their evolution is critical to many key
unsolved problems in astrophysics. It connects directly to our understanding
of star formation, as massive clusters are thought to be associated
with major star formation episodes, tracing the star-formation histories
of their host galaxies. Furthermore, globular clusters can trace the
evolution of galaxies over a significant cosmological time span, as
they are the brightest structures with which one can trace the halo
potential out to the largest radii, and they are very old, potentially
even predating the formation of their own host galaxies. Unlike stars
and planetary nebulae, globular clusters are not simply passive tracers
of galaxy kinematics as their internal dynamics are affected by the
galactic tidal field. Therefore, their internal properties and correlations
with their host galaxies are likely to contain information on the merger
history of galaxies and haloes. 

Dynamical interactions in dense star clusters play a key role in the
formation of many of the most interesting and exotic astronomical
sources, such as bright X-ray and gamma-ray sources, radio pulsars, and
supernovae. The extreme local stellar densities, which can reach $\gtrsim
10^{6}\,{\rm pc^{-3}}$,  give rise to complex dynamical processes: resonant
stellar encounters, tidal captures, physical collisions, and high-speed
ejections (\citealt{2003gmbp.book.....H}).  The primary challenge in modeling
dense clusters lies in the tight coupling of these processes and their scales
as they influence and inform one another both locally, e.g., through close
encounters or collisions on scales of $\sim 1-100\,{\rm R_{\odot}}$, or
$10^{-8}-10^{-6}$ pc, and globally on the scale of the whole system through
long-range, gravitational interactions. Close binary interactions can occur
frequently, every $\sim 10^{6}-10^{9}$ yr depending on the cluster density,
relative to the global cluster evolution timescale. Furthermore, in the time
between close encounters, stellar particles, single and binary, change their
physical properties due to their internal nuclear evolution and due to mass and
angular momentum transfer or losses. All these changes affect the rates of
close encounters and couple to the overall evolution of the cluster.

Given these enormous ranges in spatial and temporal scales, simulating dense
star clusters with a million stars or more is a formidable computational
challenge. A thorough analysis of the scaling of the computational cost of
direct $N$-body methods is presented in \citet{1988Natur.336...31H}. Although
direct $N$-body methods are free of any approximations in the stellar dynamics,
their steep $\propto N^{3}$ scaling has limited simulations to an initial
$N\sim10^{5}$ stars
(\citealp{2011MNRAS.411.1989Z,2012A&A...538A..19J,2012arXiv1208.4880H}).
However, the number of stars in real systems like globular clusters and
galactic nuclei can be orders of magnitude larger. Even for globular
  cluster size systems where the evolution is feasible to calculate with a
  direct $N$-body code, the total runtime "takes the better half of a year"
  \citep{2012arXiv1208.4880H} and statistical results have to rely on only a
  very few models. This is a problem, given the significant inherent
  stochasticity of these systems, which affects even basic structural
  parameters
\citep[e.g.,][]{2009MNRAS.397L..46H,2010ApJ...708.1598T,2012arXiv1208.4880H}.
In order to draw statistically robust conclusions, a much larger number of
realizations of massive star clusters has to be calculated, in addition to a
wider range of initial conditions. It is clear that these requirements result
in prohibitive runtimes for direct $N$-body codes.

Monte Carlo methods calculate the dynamical evolution of the cluster  in  the
Fokker-Planck approximation, which applies when the evolution of the cluster is
dominated by two-body relaxation, and the relaxation time is much larger than
the dynamical time. In practice, further assumptions of spherical symmetry and
dynamical equilibrium have to be made.  The H\'enon Monte Carlo (MC) technique
\citep{1971Ap&SS..14..151H} which is based on orbit averaging, represents a
balanced compromise between realism and speed. The MC method allows for a
star-by-star realization of the cluster, with its $N$ particles representing
the $N$ stars in the cluster. Integration is done on the relaxation timescale,
and the total computational cost scales as $N\log N$
\citep{1971Ap&SS..14..151H}.

Our work here is based on the H\'enon-type MC cluster evolution code CMC
(\textquotedblleft{}Cluster Monte Carlo\textquotedblright{}), developed over
many years by
\citet{2000ApJ...540..969J,2001ApJ...550..691J,2003ApJ...593..772F,2007ApJ...658.1047F,2010ApJ...719..915C,2012ApJ...750...31U}.
CMC includes a detailed treatment of strong binary star interactions and
physical stellar collisions \citep{2007ApJ...658.1047F}, as well as an
implementation of single and binary star evolution \citep{2010ApJ...719..915C}
and the capability of handling the dynamics around a central massive black hole
\citep{2012ApJ...750...31U}.

In addition to CMC, other MC codes have been developed recently that are
  based on the same orbit averaging technique. Apart from differences in how
  the stellar and binary process have been implemented, these codes mainly
  differ in how particles are advanced in time. The code of
  \citet{2001A&A...375..711F} uses an individual timestep scheme, where each
  particle is advanced on its own, local relaxation timescale, while the code
  of \citet[][with its newest version described in
  \citet{2011arXiv1112.6246G}]{1998MNRAS.298.1239G} uses a block timestep
  scheme, where the cluster is divided into several radial zones, and particles
  are evolved on the average relaxation timescale of the corresponding zone.
  While they provide better adaptability to density contrasts, individual and
block timestep schemes are more difficult to parallelize efficiently in a
distributed fashion. A shared timestep scheme, as implemented in CMC, offers a
greater degree of inherent parallelism \citep{2001ApJ...550..691J}.

A typical simulation starting with $N\sim 10^6$ up to average cluster ages of
12 Gyr using CMC can be run on a modern desktop computer in a reasonable amount
of time (days to weeks). However, given the scaling of computational cost,
simulations of clusters with $N\gtrsim 10^{7}$ stars, e.g., nuclear star
clusters or galactic nuclei, will still take a prohibitive amount of time.
Scaling up to even larger number of stars becomes possible only through
parallelization.

In this paper, we present in detail the latest version of CMC, which
is capable of simulating collisional systems of up to $N\sim10^{7}$.
In Section \ref{sec:Code-Overview}, we take a look at the components
of the serial code and summarize both its numerical and computational aspects.
In Section \ref{sec:Parallel-CMC}, we describe the flow of the parallel
code, elucidating how we designed each part to achieve optimal performance
on distributed parallel architectures. In addition, we describe in
the Appendix, an optional CMC feature that accelerates parts of the
code using a general purpose Graphics Processing Unit (GPU). We show
a comparison of results and analyze the performance of the code in
Section \ref{sec:Results}. Conclusions and lines of future work are
discussed in Section \ref{sec:Conclusions}.

\section{Code Overview\label{sec:Code-Overview}}

\subsection{Numerical Methods}

\begin{figure}
\includegraphics[scale=0.3]{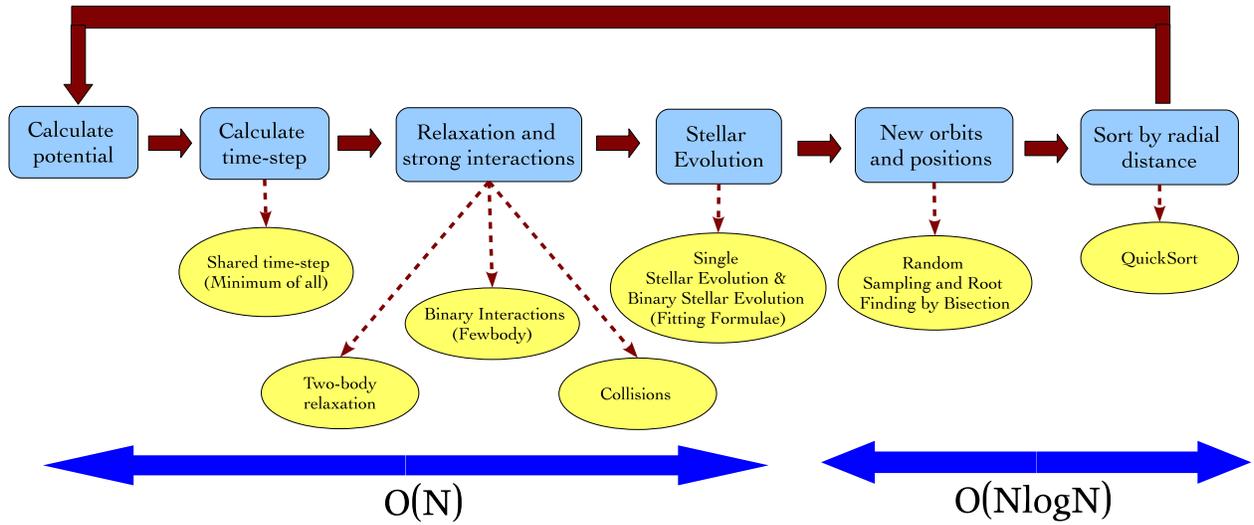}

\caption{\label{fig:cmc_flowchart}A flowchart of the CMC (Cluster Monte Carlo)
code with the following steps. (1) Potential Calculation---calculates
the spherically symmetric potential. (2) Timestep Calculation---computes
a shared timestep used by all processes. (3) Relaxation and Strong
interactions---depending on the physical system type, performs two-body
relaxation, strong interaction, or physical collision on every pair
of stars. (4) Stellar Evolution---evolves each star and binary using
fitting formulae (5) New Positions and Orbits-samples new positions
and orbits for stars. (6) Sorting---sorts stars by radial position.}
\end{figure}

Starting with an initial spherical system of $N$ stars in dynamical
equilibrium, we begin by assigning to each star, a mass, a position and
a velocity (radial and transverse components) by sampling from a distribution
function $f(E,J),$ where $E$ and $J$ are the orbital energy and
angular momentum \citep[e.g.,][]{2008gady.book.....B}. We assign
positions to the stars in a monotonically increasing fashion, so the
stars are sorted by their radial position initially. The system is
assumed to be spherically symmetric and hence we ignore the direction
of the position vector and transverse velocity. Following initialization,
the serial algorithm goes through the following sequence of steps iteratively
over a specified number of timesteps. Figure \ref{fig:cmc_flowchart}
shows the flowchart of our basic algorithm.

\begin{enumerate}
\item \emph{Potential} \emph{calculation}. The stars having been sorted by increasing radial positions in the cluster, the potential at a radius $r$, which lies between two stars at positions $r_k$ and $r_{k+1}$, is given by \begin{equation} \Phi(r) = G\left(-\frac{1}{r}\sum_{i=1}^{k} m_i - \sum_{i=k+1}^{N} \frac{m_i}{r_i}\right)\ .\\ \end{equation} where $m_i$ is the mass, and $r_i$, the position of star $i$. It is sufficient to compute and store the potential $\Phi_{k} = \Phi(r_{k})$ at radial distances $r_{k}$ $(k = 1,...,N)$, i.e., at the positions of all stars. This can be done recursively as follows::
\begin{equation}
\Phi_{N+1} = 0\ ,
\end{equation}
\begin{equation}
M_N = \sum_{i=1}^{N}m_i\ ,
\end{equation}
\begin{equation}
\Phi_k = \Phi_{k+1} - GM_k\left(\frac{1}{r_k} - \frac{1}{r_{k+1}}\right)\ ,
\end{equation}
\begin{equation}
M_{k-1} = M_{k} - m_{k}\ .
\end{equation}
To get the potential $\Phi(r)$ at any other radius, one first finds $k$ such that $r_k \leq r \leq r_{k+1}$ and then computes: \begin{equation} \Phi(r) = \Phi_{k} + \frac{1/r_{k} - 1/r}{1/r_{k} - 1/r_{k+1}} (\Phi_{k+1} - \Phi_{k})\ . \\ \end{equation}
\item \emph{Time} \emph{step} \emph{calculation}. Different physical processes
are resolved on different timescales. We use a shared timestep scheme
where timesteps for all the physical processes to be simulated are
calculated and their minimum is chosen as the global timestep for
the current iteration. The timesteps are calculated using the following
expressions \citep[see][for more details]{2007ApJ...658.1047F,2011arXiv1105.5884G,2010ApJ...719..915C}:\\
\begin{eqnarray}
T_{{\rm rel}} & = & \frac{{\theta_{max}}}{\pi/2}\frac{\pi}{32}\frac{\left<v_{rel}\right>^{3}}{\ln(\gamma N)G^{2}n\left<\left(M_{1}+M_{2}\right)^{2}\right>},\label{eq:trel}
\end{eqnarray}
\begin{equation}
T_{{\rm coll}}^{-1}=16\sqrt{\pi}n_{s}\left<R^{2}\right>\sigma\left(1+\frac{G\left<MR\right>}{2\sigma^{2}\left<R^{2}\right>}\right),\label{eq:tcoll}
\end{equation}
\begin{equation}
T_{{\rm bb}}^{-1}=16\sqrt{\pi}n_{b}X_{bb}^{2}\left<a^{2}\right>\sigma\left(1+\frac{G\left<Ma\right>}{2\sigma^{2}X_{bb}\left<a^{2}\right>}\right),\label{eq:tbb}
\end{equation}
\begin{equation}
T_{{\rm bs}}^{-1}=4\sqrt{\pi}n_{s}X_{bs}^{2}\left<a^{2}\right>\sigma\left(1+\frac{G\left<M\right>\left<a\right>}{\sigma^{2}X_{bs}\left<a^{2}\right>}\right).\label{eq:tbs}
\end{equation}
\begin{equation} T_{{\rm
  se}}=0.001M\left(\frac{T_{prev}}{\Delta{m}_{se}}\right)\ .\label{eq:tbse}
\end{equation}

where $T_{{\rm rel}},T_{{\rm coll}},T_{{\rm bb}},T_{{\rm bs}}$ and $T_{{\rm se}}$
are the relaxation, collision, binary-binary, binary-single and stellar evolution timesteps respectively. Here $\theta_{max}$ is the maximum angle of deflection
of the two stars undergoing a representative two-body encounter ;
$v_{rel}$ their relative velocities, and $n$ the local number density
of stars; $n_{s}$ and $n_{b}$ are the number densities of single
and binary stars, respectively; $\sigma$ is the one-dimensional velocity
dispersion, and $a$ is the semi-major axis. $X_{bb}$ and $X_{bs}$
are parameters that determine the minimum closeness of an interaction
to be considered a strong interaction. $M$ is the total mass of the cluster, $T_{prev}$, the previous timestep, and $\Delta{m}_{se}$, the mass loss due to stellar evolution.

The value of $T_{rel}$ is calculated for each star and the
minimum is taken as the value of the global relaxation timestep.
We use sliding averages over the neighboring 10 stars on each side
to calculate the mean quantities shown in $<\ldots>$ in Equation
\ref{eq:trel}. The other three timesteps, $T_{{\rm coll}},T_{{\rm bb}}$
and $T_{{\rm bs}}$ are averaged over the central 300 stars as in \citet{2011arXiv1105.5884G}. These choices provide a good compromise between accuracy and computational speed. Once these five timesteps are calculated, the smallest one is chosen as the timestep for the current iteration.
\item \emph{Relaxation and strong interactions (dynamics)}. Depending on the physical
system type, one of the following three operations is performed on each pair
of stars (i) Two-body relaxation is applied based on an analytic expression
for a representative encounter between two nearest-neighbor stars.
(ii) A binary interaction (binary-binary or binary-single) is simulated using
\texttt {Fewbody}, an efficient computational toolkit for evolving small-$N$
dynamical systems (\citealt{2004MNRAS.352....1F}). {\texttt Fewbody} performs a
direct integration of Newton\textquoteright{}s equations for 3 or 4 bodies
using the $8$th-order Runge-Kutta Prince-Dormand method.  (iii) A stellar
collision is treated in the simple ``sticky sphere'' approximation, where two
bodies are merged whenever their radii touch, and all properties are changed
correspondingly (\citealt{2007ApJ...658.1047F}).  \item \emph{Stellar
  Evolution}. We use the SSE (\citealt{2000MNRAS.315..543H}) and BSE
  (\citealt{2002MNRAS.329..897H}) stellar evolution routines, which are based
  on analytic functional fits to theoretically calculated stellar evolution
tracks, to simulate the evolution of single and binary stars
(\citealt{2010ApJ...719..915C}).  \item \emph{New orbits computation}.
  Consistent with the changes in the orbital properties of the stars following
  the interactions they undergo, new positions and velocities are assigned for
  orbits according to the new energy and angular momentum. Then, a new position
  is randomly sampled according to the amount of time the star spends near any
  given point along the orbit. Since this step represents a major computational
  bottleneck, we provide some details here.

We start by finding the pericenter and apocenter distances of the star's new orbit.  Given a star with specific energy $E$ and angular momentum $J$ moving in the gravitational potential $\Phi(r)$, its rosette-like orbit $r(t)$ oscillates between two extreme values $r_\text{min}$ and $r_\text{max}$, which are roots of: 
\begin{equation} 
Q(r) = 2E - 2\Phi(r) - J^2/r^2 = 0\ .
\end{equation} 
Since we only store the potential values at the positions of the stars, this equation cannot be analytically solved  before determining the interval in which the root lies. In other words, we need to determine $k$ such that $Q(r_k) < 0 < Q(r_{k+1})$. We use the bisection method to do this. Once the interval is found, $\Phi$, and thus $Q$, can be computed analytically in that interval, and so can $r_\text{min}$ and $r_\text{max}$.

The next step is to select a new radial position for the star between $r_\text{max}$ and $r_\text{min}$. The probability is equal to the fraction of time spent by the star in $dr$, i.e.:
\begin{equation} 
f(r) = \frac{dt}{T} = \frac{dr/\left|{v_\text{r}}\right|}{\int_{r_\text{min}}^{r_\text{max}} {dr/\left|{v_\text{r}}\right|}} \ ;
\end{equation} 
where the radial velocity $v_r = [Q(r)]^{1/2}$. We use the von Neumann rejection technique to sample a position according to this probability. We take a number $F$ which is everywhere larger than the probability distribution $f(r)$. Then we draw two random numbers $X$ and $X'$ and compute
\begin{equation} 
r_0 =  r_\text{min} + (r_\text{max} - r_\text{min}) X\ , \end{equation} 
\begin{equation}
f_0 = FX'\ .
\end{equation}
If the point $(f_0,r_0)$ lies  below the curve, i.e., if $f_0 < f(r_0)$, we take $r=r_0$ as the new position; otherwise we reject it and draw a new point in the rectangle with a new pair of random numbers. We repeat this until a point below the curve is obtained. In our code, a slightly modified version of the method is used, since $f(r) = 1/|v_r|$ becomes infinite at both ends of the interval. A detailed description can be found in \citet{2000ApJ...540..969J}.
\item \emph{Sort stars by radial distance.} This step uses the Quicksort
algorithm (\citealp{Hoare:1961:AQ:366622.366644}) to sort the stars
based on their new positions. Sorting the stars is essential to determine
the nearest neighbors for relaxation and strong interactions, and
also for computing the gravitational potential.
\item \emph{Diagnostics, energy conservation, and program termination}. These involve the computation of diagnostic values such as half-mass radius, core radius, number of core stars etc., that control program termination. In addition, corrections are made to
    the kinetic energy of each particle that account for the fact that the new
    orbit has been calculated in the old potential from the previous timestep.
    This is done by adjusting the stellar velocities according to the
    mechanical work done on the stellar orbit by the time varying potential. We
    mainly follow the procedure in \citet{1982AcA....32...63S} except that we
    apply the correction to the velocity components such that the ratio
  $v_r/v_t$ is preserved. See \citet{2007ApJ...658.1047F} for more details.
 Since these minor calculations are scattered throughout various places in the code, they are not explicitly shown in the flowchart (Figure \ref{fig:cmc_flowchart}).
\end{enumerate}

\subsection{Time Complexity Analysis}

In addition to the flow of the code, Figure \ref{fig:cmc_flowchart}
also shows the time complexity for each of the above steps. The timesteps, $T_{{\rm coll}},T_{{\rm bb}}$ and $T_{{\rm bs}}$ are averaged over a fixed number of central stars, whereas $T_{{\rm se}}$ is a simple factor of the previous timestep, and hence are ${\cal O}(1)$ operations.  The calculation
of $T_{{\rm rel}}$ for a single star involves averaging over a fixed number of neighboring
stars and hence has constant time complexity, ${\cal O}(1)$. As this is done for all $N$ stars
to estimate their individual timesteps from which the  the minimum is chosen, the timestep calculation scales as ${\cal O}(N)$. The effect of relaxation and strong interactions
 is calculated between pairs of stars that are radially nearest neighbors.
Since these  calculations involve constant time operations for each
of the $N$ stars, the time complexity of the perturbation step is ${\cal O}(N)$.
Stellar evolution operates on a star-by-star basis performing operations
of constant time for a given mass spectrum, and hence also scales
as ${\cal O}(N)$. Determination of new orbits for each star involves
finding the roots of an expression on an unstructured one-dimensional
grid using the bisection method. The bisection method scales as ${\cal O}(\log N)$ and
as this is done for each star, this step has a time complexity of
${\cal O}(N\log N)$. The radial sorting of stars using Quicksort
has the same time complexity \citep{Hoare:1961:AQ:366622.366644}.

\section{Parallelization\label{sec:Parallel-CMC}}

Achieving optimal performance of codes on parallel machines require serial algorithms
to be carefully redesigned, and hence, parallel algorithms are often
very different than their serial counterparts and require a considerable
effort to develop. The key to a successful algorithm is (1) good load
balance, i.e., the efficient utilization of the available processing
units, and (2) minimal communication between these units. The communication
cost depends directly on the choice of domain decomposition, i.e,
the way in which work and data is partitioned into smaller units
for processing in parallel. For example, a good domain decomposition
scheme for achieving ideal load balance would be the distribution of stars (i.e., their data)
evenly among the processors, assuming the computational cost for processing
each star is similar. This will ensure the workload is evenly balanced
across processors given that they all perform the same number of operations,
as in a Single Program, Multiple Data (SPMD) programming model. However,
how such a decomposition would influence the need for communication
between processing units is very specific to the algorithms used. In essence,
a thorough knowledge of the algorithm, and its data access patterns
is necessary for designing any efficient parallel application.

\subsection{Data Dependencies and Parallel Processing Considerations}

While deciding upon the domain decomposition, we have to take into
account any dependencies, i.e., the way the data is accessed by various
parts of the application, as they may directly influence both the
load balance and the communication costs between the processing units.
A good parallel algorithm should distribute the workload in such a
way that the right balance is struck between load balance and communication,
resulting in optimal performance. 

In CMC, the physical data of each star (mass, angular momentum, position
etc.) are stored in a structure, a grouping of data elements. The
data for $N$ stars are stored an array of such structures. For a system with
$p$ processors and $N$ initial stars, we will first consider the
scenario where we try to achieve ideal load balance by naively distributing
the data of $N/p$ stars to each processor. Fo simplicity, we will assume here that
$N$ is divisible by $p$, and analyze the data dependencies
in the various modules of CMC for this decomposition.
\begin{enumerate}
\item \emph{Timestep Calculation}:

\begin{enumerate}
\item For calculating the relaxation time of each star we need the local
density, which is calculated using the masses of the 10 nearest neighboring
stars on either side of the radially sorted list of stars. A parallel
program would require communication between processors to exchange data
of the neighboring stars that are at the extreme ends of the local
array. 
\item Calculation of the timestep requires the computation of quantities
in the innermost regions of the cluster, in particular the central
density, and the half-mass radius. If the particles are distributed
across many processors, irrespective of the specific data partitioning
scheme, identification of the particle up to which the enclosing stellar
mass equals half the total mass of the cluster might require communication
of intermediate results between adjacent data partitions, and also
would introduce an inherent sequentiality in the code.
\end{enumerate}
\item \emph{Relaxation and strong interactions (dynamics):} \\
For the perturbation calculation, pairs of neighboring stars  are
chosen. Communication might be needed depending on whether the number
of stars in a processor is even or odd.
\item \emph{New} \emph{orbits} \emph{computation}: \\
To determine the new orbits of each star we use the bisection method
which involves random accesses to the gravitational potential profile
of the cluster. Since this data will be distributed in a parallel algorithm,
communication will be needed for data accesses that fall outside the
local subset.
\item \emph{Sorting}: \\
Putting the stars in order according to their radial positions naturally
needs communication irrespective of the decomposition.
\item \emph{Potential Calculation}: The potential calculation, as explained
in Section \ref{sec:Code-Overview}, is inherently sequential and requires
communication of intermediate results.

\item \emph{Diagnostics and program termination:} \\ 
  The diagnostic quantities that are computed on different computational units
  need to be aggregated before the end of the timestep to check for errors or
  termination conditions.  
  
\end{enumerate}

\subsection{Domain Decomposition and Algorithm Design\label{sub:Domain-Decomposition}}

Based on the considerations in the previous section, we design the
algorithms and decompose the data according to the following scheme
so as to minimize communication costs, and at the same time not degrading
the accuracy of the results.

Since the relaxation timestep calculation procedure introduces additional
communication cost irrespective of the choice of data partitioning,
we modify it in the following way. Instead of using sliding averages
for each star, we choose radial bins containing a fixed number of
stars to calculate the average quantities needed for the timestep
calculation. We choose a bin size of 20 which gives a good trade off
between computational speed and accuracy. We tested this new timestep
calculation scheme, and we did not find any significant differences
compared to the previous scheme. In addition, we distribute the stars
such that the number of stars per processor is a multiple of 20 (for
the timestep and density estimates) for the first $(p-1)$ processors
and the rest to the last processor. Since 2 is a multiple of 20, this
removes any dependencies due to the interactions part as well. Our choice
of data partitioning also ensures a good load balance as, in the worst
case, the difference between the maximum and minimum number of stars
among the processors could be at most 19.

The gravitational potential $\Phi(r)$ is accessed in a complex,
data dependent  manner as we use the bisection method to determine
new stellar orbits. Hence, we do not partition it among the processors,
but maintain a copy of it on all nodes. We also do the same for the
stellar masses and positions,  to remove the dependency in the potential calculation.
This eliminates the communication required by the new orbits and potential
calculations. However, it introduces the requirement to keep these
data arrays synchronized at each timestep and hence adds to the communication.
We estimate the communication required for synchronization to be much
less than what would be added by the associated dependencies without
the duplicated arrays.

Most modules of the code perform computations in loops over the local
subset of stars that have been assigned to the processor. Depending
on the computation, each processor might need to access data from
the local and duplicated arrays. While the local arrays can be  accessed
simply using the loop index, any accesses of the duplicated arrays
(potential, position, or mass) require an index transformation. For
instance, let us consider a simple energy calculation routine that
calculates the energy of each star in the local array over a loop
using the equation
\begin{equation}
E_{i}=\Phi_{gi}+0.5\,(v_{r,i}^{2}+v_{t,i}^{2})\:.
\end{equation}

where $E_{i},v_{r,i}$ and $v_{t,i}$ are the energy, radial and transverse
velocities of star $i$ in the local array. The potential array having
been duplicated across all processors, the potential at the position
of the $i$th star is $\Phi_{gi}$, where $gi$ is the global index
given by the following transformation which directly follows from
our data partitioning scheme explained above:
\begin{equation}
gi=\begin{cases}
i+id\,\left\lfloor \left\lfloor \frac{N}{n_{m}}\right\rfloor \,\frac{1}{p}\right\rfloor \, n_{m}+id\, n_{m} & \mbox{for }id<\left\lfloor \frac{N}{n_{m}}\right\rfloor \bmod{p}\ ,\\
i+id\,\left\lfloor \left\lfloor \frac{N}{n_{m}}\right\rfloor \,\frac{1}{p}\right\rfloor \, n_{m}+\left\lfloor \frac{N}{n_{m}}\right\rfloor \bmod{p}\, n_{m} & \mbox{for }id\geq\left\lfloor \frac{N}{n_{m}}\right\rfloor \bmod{p}\ .
\end{cases}
\end{equation}

where $id$ the id of the processor that is executing the current
piece of code, which ranges between 0 to $p-1$, $n_{m}$ is the number
we would want the number of stars in each processor to be
a multiple of, which as per our choice, is 20, and the terms between
$\left\lfloor \ldots\right\rfloor $ are rounded to the lowest integer.

\subsection{Parallel Flow}

The following gives an overview of the parallel workflow:
\begin{enumerate}
\item Initial partitioning of the star data and distribution of duplicated
arrays (mass, and radial positions)
\item Calculate potential
\item Perform interactions, stellar evolution, and new orbits calculation
\item Sort stars by radial position in parallel
\item Redistribute/load-balance data according to domain decomposition
\item Synchronize duplicated arrays (mass, and radial positions)
\end{enumerate}
Then the whole procedure is repeated starting from step 2. The first
step is to distribute the initial data among processors as per our
data partitioning scheme mentioned in Section \ref{sub:Domain-Decomposition}.
This is done only once per simulation. This also includes the distribution
of a copy of the duplicated arrays. In Section \ref{sec:Code-Overview},
we saw that the calculation of the potential is inherently sequential
requiring communication, since it is calculated recursively starting
from the outermost star and using the intermediate result to compute
the potential of the inner stars. However, since now every processor
has an entire copy of the potential, the positions and mass arrays,
it can calculate the potential independently. This does not give a
performance gain since there is no division of workload, however nullifies
the communication requirement. Performing interactions, stellar evolution
and new orbits calculation too don't require any communication whatsoever
due to our choice of data partitioning and use of duplicated arrays.
We use {\texttt Sample Sort} (\citealt{samplesort}) as the parallel sorting algorithm (see Section \ref{sub:Sorting}). With a wise choice
of parameters, {\texttt Sample Sort} can provide a near equal
distribution of particles among processors. However, since we require
the data to be partitioned in a very specific way, following the sort,
we employ a redistribution/load-balancing phase to redistribute the
sorted data as per our chosen domain decomposition. Sorting and redistribution
are steps that naturally require the most communication. Before the
beginning of the next timestep, we synchronize the duplicated arrays
on all processors which requires message passing. Some non-trivial
communication is also required at various places in the code to collect
and synchronize diagnostic values.

\subsection{Sorting\label{sub:Sorting}}

The input to a parallel sorting algorithm consists of a set of data
elements (properties of $N$ stars in our case), each having a key
(radial positions in our case) based on which the data need to be
sorted. An efficient parallel sorting algorithm should collectively
sort data owned by individual processors in such a way that their
utilization is maximum, and at the same time the cost of redistribution
of data across processors is kept to a minimum. In order to implement
a parallel sorting algorithm, there are a wide array of solutions
to consider. Each of these solutions cater to a parallel application
and/or in some cases a particular machine architecture/platform. In
general, a parallel programmer should consider many different design
decisions with careful consideration of the application characteristics.
The proper assessment of application knowledge often can suggest which
initial distributions of data among processors are likely to occur,
allowing the programmer to implement a sorting algorithm that works
effectively for that application.

The importance of load balance is also immense, since the application's
execution time is typically bound by the local execution time of the
most overloaded processor. Since our choice of domain decomposition
requires a fairly good load balance, our sorting algorithm should
ensure that the final distribution of keys among processors closely
agree with our decomposition. This is a challenge since during their
evolution, dense star clusters have a very strong density contrast,
and stars are very unevenly distributed radially with a substantial
number of stars in the high density core, and a the rest in the extremely
low density halo. A good parallel sorting algorithm for our application
should be able to judge the distribution of keys so as to deliver almost equal amount of data in each processor
at the end of the sorting phase.

\begin{figure}
\includegraphics[scale=0.65]{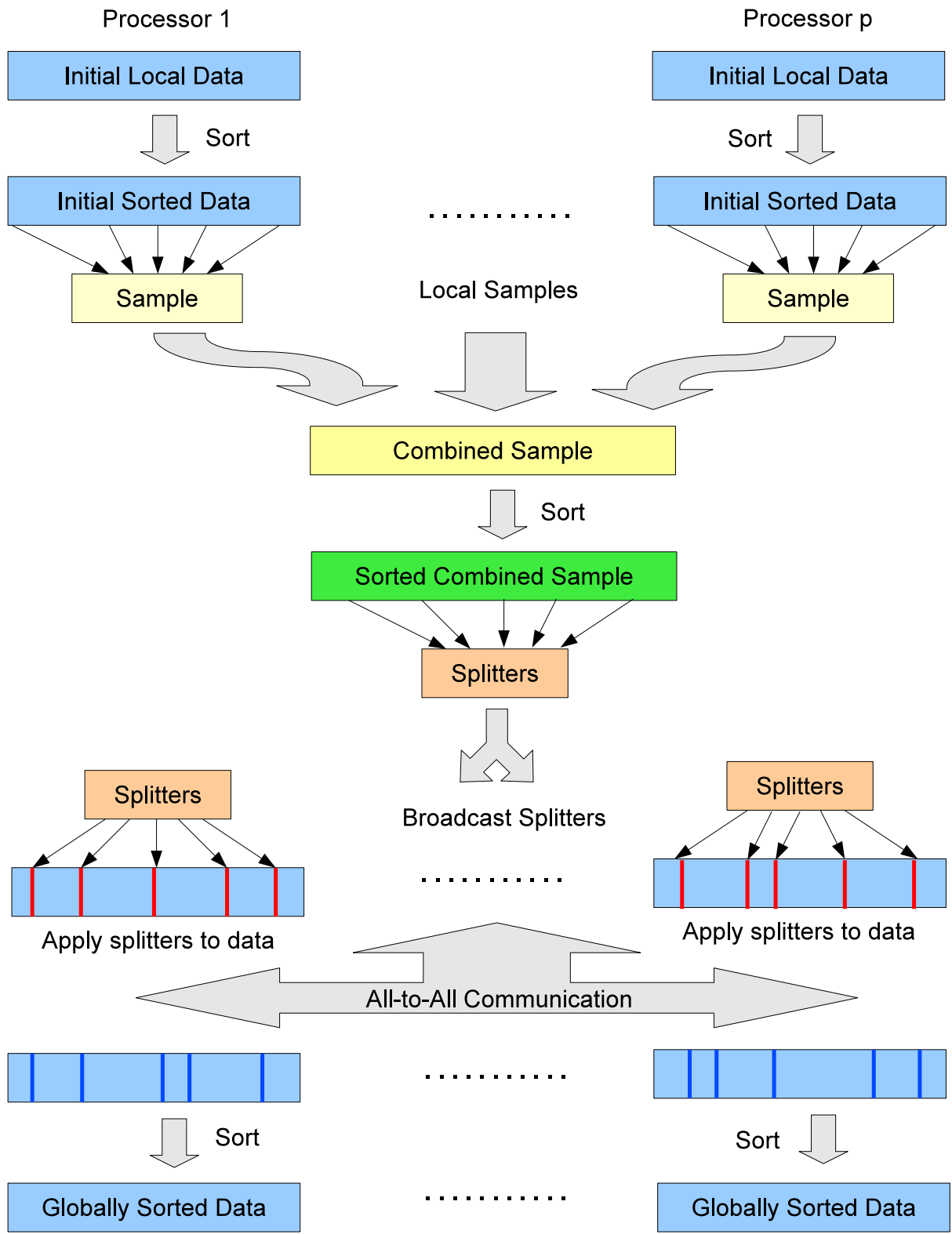}

\caption{\label{fig:The-Sample-Sort}The {\texttt Sample Sort} Algorithm}

\end{figure}

{\texttt Sample Sort} is a splitter-based parallel sorting algorithm that performs
a load balanced splitting of the data by sampling the global key set.
This sampling helps judge the initial distribution of keys and accordingly
perform the sort, hence resulting in a near-equal distribution of
data among processors. Given $N$ data elements distributed across
$p$ processors, {\texttt Sample Sort} consists of 5 phases, shown in Figure
\ref{fig:The-Sample-Sort}:
\begin{enumerate}
\item \textbf{Sort Local Data}: Each processor has a contiguous block of
data in accordance with our data partition (close to $N/p$ in number,
see Section \ref{sub:Domain-Decomposition}). Each processor, in parallel,
sorts this local block using sequential Quicksort.
\item \textbf{Sampling}: All $p$ processors, in parallel, uniformly sample
$s$ keys to form a representative sample of the locally sorted block
of data. These set of $p$ samples, of size $s$ from each processor,
are collected on one designated processor. This aggregated array of
samples represent the distribution of the entire set of keys.
\item \textbf{Splitter Selection}: The combined sample key array is sorted,
and keys at indices $s,2s,3s,...,(p-1)s$ are selected as splitters
and are broadcasted to all processors. 
\item \textbf{Exchange partitions}: The positions of the $(p-1)$ splitter
points in the local array are determined by each processor using binary
search; this splits the local array into $p$ partitions. In parallel,
each processor retains the $i$th partition and sends the $j$th partition
to the $j$th processor, i.e. each processor keeps 1 partition and
distributes $(p-1)$ partitions. At the same time it receives 1 partition
from every other processor. This might not be true always, particularly
in cases of a poor choice of sample size $s$, some splitter points
might lie outside the local data array and hence some processors might
not send data to all $(p-1)$ processors but only a subset of them.
However, for the current discussion we will assume a good choice of
sample size is made.
\item \textbf{Merge Partitions}: Each processor, in parallel, merges its
$p$ partitions into a single list and sorts it.
\end{enumerate}
One chooses a value for the sample size
$s$ so as to sample the distribution of keys accurately, and hence
this parameter varies depending on the distribution of keys as well
the data size $N$. More comments on the choice of sample size can
be found in \citet{Li19931079}.

Let us now try to derive the time complexity of {\texttt Sample Sort}. We assume a hypercube
parallel architecture with cut-through routing, a message routing mechanism in which nodes immediately forward messages to subsequent ones as they arrive\footnote{This is faster than when the nodes wait until the entire message is received and stored before forwarding, also known as store-and-forward routing.}. The local sort (Phase 1) requires
${\cal O}(\frac{N}{p}\log(\frac{N}{p}))$ since there are close to
$N/p$ keys per processor. The selection of $s$ sample keys (Phase 2, part 1)  requires
${\cal O}(s)$ time. Collecting $s$ keys from $p$ processors on
to one of the processors (Phase 2, part 2) is a single-node gather operation for which
the time required is ${\cal O}(sp)$. The time to sort these $sp$
samples is ${\cal O}((sp)\log(sp))$, and the time to select $(p-1$)
splitters (Phase 3, part 1) is ${\cal O}(p)$. The splitters are sent to all the other
processors (Phase 3, part 2) using an one-to-all broadcast which takes ${\cal O}(p\log p)$
time. To place the splitters in the local array (Phase 4, part 1) using binary search
takes ${\cal O}(p\log(\frac{N}{p}))$. The all-to-all communication
that follows (Phase 4, part 2) costs a worst case time of ${\cal O}(\frac{N}{p})+{\cal O}(p\log p)$. The final step that merges and sorts the partitions would cost ${\cal O}(\frac{N}{p}\log(\frac{N}{p}))$ time.
So the time complexity of the entire algorithm becomes

\begin{equation}
{\cal O}\left(\frac{{N}}{p}\log\frac{{N}}{p}\right)+{\cal O}((sp)\log(sp))+{\cal O}\left(p\log\frac{{N}}{p}\right)+{\cal O}(N/p)+{\cal O}(p\log p)\ .\label{eq:sort_complexity}
\end{equation}

\subsection{Data Redistribution\label{sub:Data-Redistribution}}

In theory, with a good choice of sample size, {\texttt Sample Sort} guarantees
to distribute the particles evenly among processors within a factor
of two \citep{Li19931079}. However, we would like to partition the
data in such a way that each processor has close to $N/p$ elements,
and at the same time being multiple of 20. Since the final distribution
of data among processors after {\texttt Sample Sort} is not deterministic, we
include an additional communication phase to ensure the required domain decomposition is maintained.

We first calculate the global splitter points that would partition
the entire data among processors as per our required data partitioning
scheme. We then perform a parallel prefix reduction ($MPI\_Exscan$), so
each processor knows the cumulative number of stars that ended up
in the processors before it. Using this, it can calculate the range of
global indices corresponding to the list of stars it currently holds.
Now, each processor checks if any of the global splitter points other
than its own map on to its local array, and if they do, it marks the
corresponding chunk of data to be sent to the respective processor.
Then, we perform an all-to-all communication, after which the data on each processor is 
sorted by simply rearranging the received chunks.

Let us consider an example where there are 4 processors and they
receive 100, 130, 140 and 80 stars respectively after the sorting phase. For
a total of 450 stars to be divided among 4 processors, the expected
distribution would be 120, 120, 100 and 110 stars respectively as
per our scheme. The corresponding global splitter points would be
120, 240, and 340. By doing the prefix reduction on the received number
of stars, processor 3, for instance, knows there are in total 230
stars in processors 1 and 2 put together. Since it received 140 stars after sorting,
it also calculates that it has stars with indices between 231 - 370.
Now, two global splitter points, i.e., 240 and 340 of processors 2
and 4 lie within this index range, and hence the corresponding stars,
i.e., 231 - 240 and 341 - 370 need to be sent to processors 2 and
4 respectively. These data chunks are exchanged by performing an all-to-all
communication, followed by a rearrangement if required.

\subsection{Parallel Random Number Generation\label{sub:Parallel-Random-Number}}

The accuracy of results of a parallel Monte Carlo code depends in
general  on both the quality of the pseudo-random number generators
(PRNGs) used and the approach adopted to generate them in parallel.
 In our case we need to sample events with very low probability, such
as physical collisions between stars or binary interactions, which
makes it necessary for the generated random numbers to be distributed
very evenly. More specifically, given $N_{r}$ random numbers uniformly
distributed in the interval $(0,1)$, it would be preferable to have one random number
in each sub-interval of size $1/N_{r}$. A special class of random
number generators for which this property holds in even higher dimensions
are the maximally equidistributed generators  and we choose here
the popular and fast ``combined Tausworthe linear feedback shift register" (LFSR)
PRNG in \citet{L'Ecuyer:1999:TME:307090.307103}.

PRNGs use a set of state variables which are used to calculate random
numbers. Every time a random number is generated, the values of these
states are updated. A PRNG can be initialized to an arbitrary starting
state using a random seed. When initialized with the same seed, a
PRNG will always produce the same exact sequence. The maximum length
of the sequence before it begins to repeat is called the period of
the PRNG. The combined Tausworthe LFSR PRNG we are using here has a 
period of $\approx2^{113}$ \citep{L'Ecuyer:1999:TME:307090.307103}.

While generating random numbers  in parallel, special care has to
be taken to ensure statistical independence of the results calculated
on each processor. For instance, to implement a parallel version,
we could simply allocate separate state variables for the PRNGs
on each processor and initialize them with a different random seed.
However, choosing different seeds  does not guarantee statistical
independence between these streams.

An efficient way to produce multiple statistically independent streams
 is to divide  a single random sequence into subsequences, with their
starting states calculated using jump functions \citep{Army_report}.
Taking a starting seed and a jump displacement, D, as inputs, jump
functions can very quickly generate the $D$th  state of the random
sequence. We use this method to generate multiple starting states,
one for each processor, by repeatedly applying the jump function and
saving intermediate results. The jump displacement is chosen as the
maximum number of random numbers each processor might require for
the entire simulation while still providing for a sufficiently large
number of streams. Based on that  we choose $D=2^{80}$.

\subsection{Implementation}

CMC is written in C, with some parts in Fortran. We use the Message
Passing Interface (MPI) library (\citealt{mpibasic}) to handle communication. The MPI standard
is highly portable to different parallel architectures, and available
on practically every supercomputing platform in use today. The most
common MPI communication calls (see \citealt{mpibook} for a detailed description) used in our code are:
\begin{enumerate}
\item $MPI\_Allreduce/MPI\_Reduce$\\
MPI\_Reduce combines the elements of a distributed array by cumulatively
applying a user specified operation as to reduce the array to a single
value. For instance, when the operation is addition then the resulting
value is the sum of all elements. MPI\_Allreduce is MPI\_Reduce with
the difference that the result is distributed to all processors. The
call is used in the following parts of the code:

\begin{enumerate}
\item \emph{Diagnostics and program termination}: accumulating diagnostic
quantities such as the half-mass radius, $r_{h}$, and the core radius,
$r_{c}$.
\item \emph{Timestep calculation}: to find the minimum timestep of all stars
across processors.
\item \emph{Sorting and data redistribution}: Since stars are created and lost
throughout the simulation, $N$ is not a constant and  changes during
a timestep. It is calculated during the sorting step by summing up
the local number of stars on each processor.
\end{enumerate}
\item $MPI\_Allgather/MPI\_Gather$\\
In MPI\_Gather each process sends the contents of its local array
to the root, or master, process, which then concatenates all received
arrays into one. In MPI\_Allgather this concatenated array is distributed
to all nodes. The calls are used in the following parts of the code:

\begin{enumerate}
\item Synchronization of duplicated arrays, i.e., $\Phi(r)$ and the stellar
masses.
\item Sorting and data redistribution: to gather samples contributed by
all processors on to a single node. See Section \ref{sub:Sorting}
for details.
\end{enumerate}
\item $MPI\_Alltoall$\\
In MPI\_Alltoall the send and receive array is divided equally into
p sub-arrays, where p is the number of processors. The position of
each sub-arryay within the send or receive array determines to or
from which processor the data is sent or received, respectively. MPI\_Alltoall
is only used in ``Sorting and data redistribution''. See Section
\ref{sub:Sorting} for details.

\item $MPI\_Bcast$\\
In MPI\_Bcast an array is sent from the root, or master, node to all
other processes.\\
Used in ``Sorting and data redistribution'' to communicate the new
splitter points from a single specified processor to all other.
\item $MPI\_Scan/MPI\_Exscan$\\
MPI\_Scan essentially carries out a reduction as in MPI\_Allreduce
except that processor $i$ receives the result of the reduction over
the data of processors 0 to i. In MPI\_Exscan the data is reduced
over processors 0 to i-1. MPI\_Scan/Exscan is used in \emph{Sorting and
data redistribution}. See Section \ref{sub:Data-Redistribution}
for details.

\end{enumerate}
We also use a number of optimizations for the MPI communication calls.
Some examples include using MPI derived datatypes for data packing
and sending, and combining multiple parallel reduction calls for the
diagnostic quantities by packing all the data into a single buffer
and performing the reduction together using a single call which is
more efficient. However, the overlapping of communication calls with
computation we did not explore so far, but  intend to do so  in the
future.

\section{Results\label{sec:Results}}

All our test calculations were carried out on {\texttt Hopper}, a Cray XE6 supercomputer
at NERSC%
\footnote{\url{http://www.nersc.gov/}%
} that has a peak performance of 1.28 Petaflops, 153,216 processor-cores
for running scientific applications, 212 TB of memory, and 2 Petabytes
of online disk storage.

\subsection{Accuracy and Reproducibility}

In the parallel version, since we use several different random sequences
within one simulation, the way random numbers are assigned to stars is
different from the serial version. This would bring in a problem of
inconsistency in the results between serial and parallel runs, leaving us with
no simple way to verify the correctness of the results of our parallel version.
We tackle this problem by changing the serial version such that it uses the
same mapping of random number streams to stars as followed in the
parallel version. This emulation allows us to compare the results of the
  parallel version with that of the serial version. However, note that
parallel runs with different numbers of processors will result in a different
mapping of streams to stars, making a separate serial run necessary to compare
results for each case.

We ran test simulations for 50 timesteps with the parallel and serial
  versions, using or emulating up to a few processors, respectively. Initially,
  the clusters had N = $10^{5}$, $10^{6}$, and $10^{7}$ stars with positions
  and velocities distributed according to a Plummer model. By comparing the
  positions and masses of every star in the cluster at the end of the
  simulations, we found that the parallel and corresponding serial results
  were matching accurately down to the last significant digit (all variables
are in double precision). We also compared a few diagnostic quantities, 
such as the core radius, and core density, and they were matching as well, except
for the last four significant digits. This slight loss in accuracy is due to
the $MPI\_Reduce$ calls, which perform cumulative operations (sum, max, min
etc.) on data distributed among the processors. This introduces different
round-off errors since one does not have control over the order in which the
data aggregation is done.

\subsection{Comparison to Theoretical Predictions}

\begin{figure}
\includegraphics[scale=0.65]{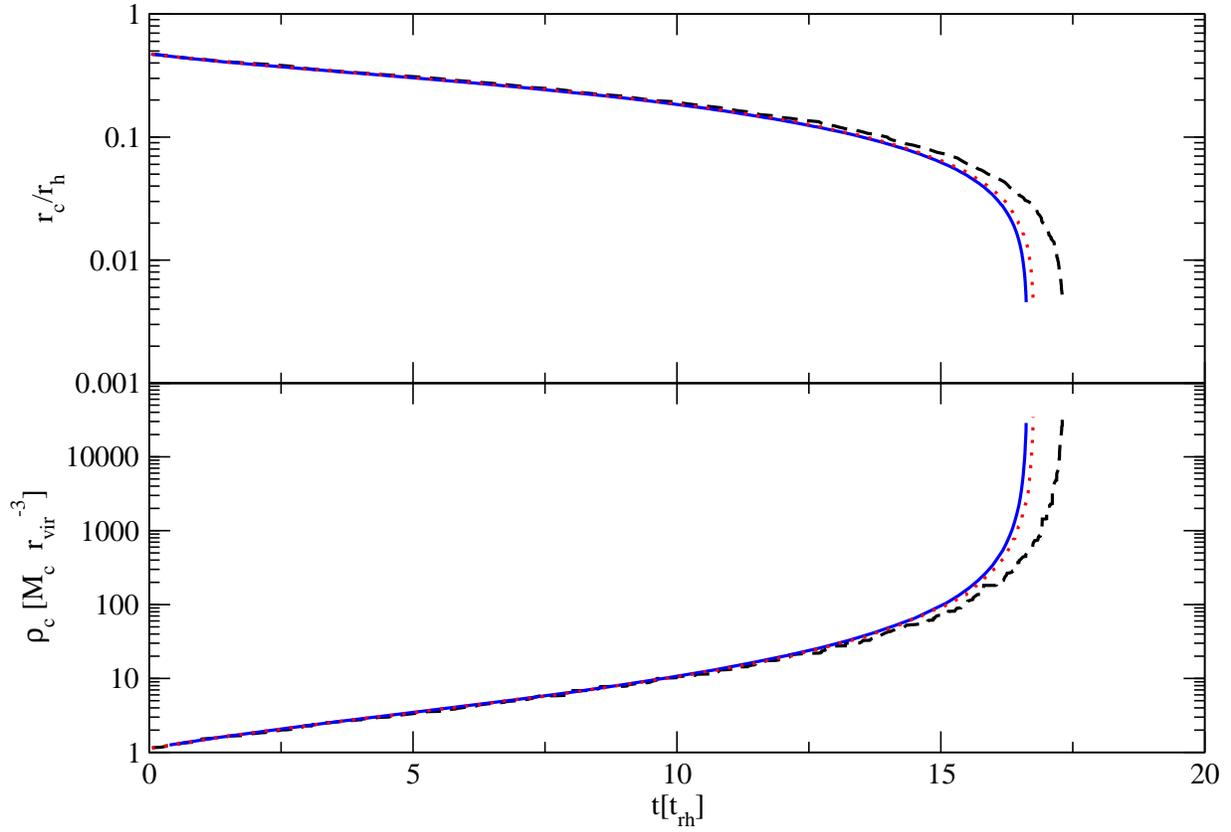}

\caption{\label{fig:rcrh}Evolution of an isolated Plummer model showing the
ratio of core radius to half-mass radius, $r_{{\rm c}}/r_{{\rm h}}$ (top), and
the core density, $\rho_{c}$ (bottom). Time is in initial half-mass relaxation
times. The various lines represent different particle numbers,
$N=10^{5}\,(\mbox{dashed}),10^{6}\,(\mbox{dotted}),\,10^{7}\,(\mbox{solid})$ .}
\end{figure}

In order to verify that our code reproduces well-known theoretical
results, we calculate the evolution of single-mass Plummer spheres
\citep{2008gady.book.....B} until core collapse (without any binaries or
stellar evolution). With $10^{5},\,10^{6},$ and $10^{7}$ stars, this is
the first time that collisional $N$-body simulations covering three orders
of magnitude up to $N=10^{7}$ stars have been carried out.  We used 128, 256
and 1024 processors for these runs, respectively, which deliver peak
performance for the three cases (see Section \ref{sub:Performance-Analysis}).
The wall clock run times for these simulations were 11.4 mins, 1.17 hrs
and 12.8 hrs, respectively.

\begin{figure}
\includegraphics[scale=0.65]{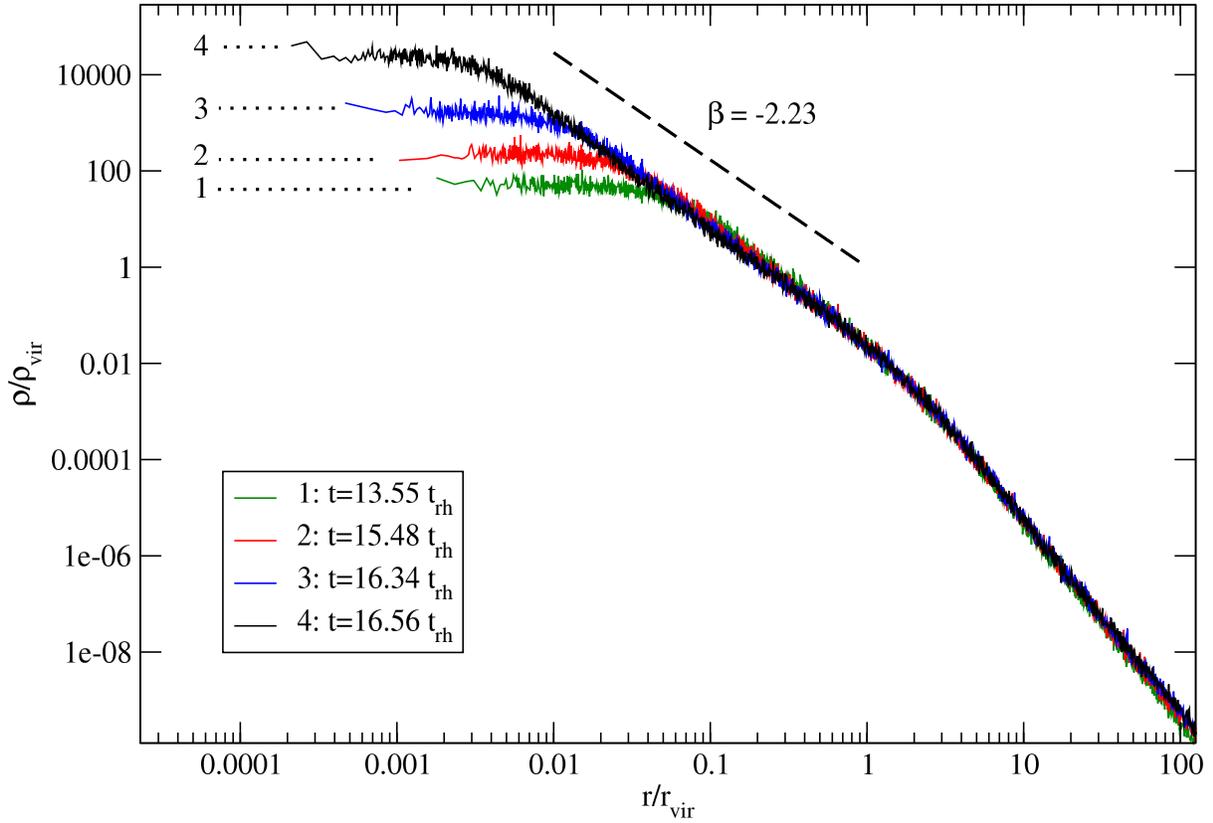}

\caption{\label{fig:density_profile}Evolution of the density profile at various
times during core collapse for the $N=10^{7}$ run. The dashed line shows the
slope of the expected power-law density profile
\citep{1988MNRAS.230..223H,1980ApJ...242..765C}.} 

\end{figure}

One remarkable property realized early on is that the cluster evolution
proceeds asymptotically in a self-similar fashion, that is, the cluster density
profiles differ only in scale and normalization at late times
\citep[e.g.,][]{1980ApJ...242..765C,2008gady.book.....B}.  This can be clearly
seen in Figure \ref{fig:density_profile}, where we plot the density profile of
the cluster with $N=10^{7}$ at various times during its evolution to core
collapse. For each profile we observe a well-defined core and a power law
density profile, $\rho\propto r^{-\beta}$, with $\beta\approx2.3$. This
is only slightly larger than the value $\beta=2.23$ usually quoted in
the literature (first obtained by \citep{1980ApJ...242..765C}). The
slight discrepancy, however, arises probably because the density profiles in Figure
\ref{fig:density_profile} were taken at a time when core collapse has not yet
proceeded far enough.

Figure \ref{fig:rcrh} shows the evolution of the core radius, $r_{{\rm
c}}(t)/r_{{\rm h}}(t)$, as well as the core density, $\rho_{c}(t)$, for the
models with $N=10^{5}$, $10^{6}$ and $10^{7}$ stars. 
We use the notation
  $r_c$ for the density-weighted core radius \citep{1985ApJ...298...80C}, and
  $\rho_c$ for the core density, defined as

\begin{equation}
\rho_c=\frac{\sum_i{\left[\rho_i\right]^2}}{\sum_i{\rho_i}}
\end{equation}

where $\rho_i$ is the 6th order density estimator around the $i$th star
\citep{1985ApJ...298...80C}. The core density $rho_c$ is expressed in units of
$M_c\,r_{vir}^{-3}$, where $M_c$ is the total cluster mass and $r_{vir}$ is the
virial radius, defined as $GM_c^2/(2E_0)$ with $E_0$ the initial total
gravitational energy of the cluster, and $G$ the constant of gravity. One can
immediately see that all three clusters reach core collapse at similar times,
with $t=t_{cc}\simeq17.4t_{{\rm rh},}16.7t_{{\rm rh}}$ and $16.6\, t_{{\rm
rh}}$, respectively, where $t_{{\rm rh}}$ is the initial half-mass
relaxation time defined as \citep{1987degc.book.....S}

\begin{equation}
t_{rh}=\frac{0.138N}{\ln(\gamma N)}\left(\frac{r_h^3}{GM}\right)^{1/2}\ .
\end{equation}

with $\gamma$ in the Coulomb logarithm chosen to be $0.1$. Thus, the core
collapse times are not only in very good agreement with previously published
results that all lie between 15 and 18 $t_{rh}$ \citep[see,~e.g.,][for an
overview]{2001A&A...375..711F}, but also confirm that our code can reproduce
  the scaling of $t_{{\rm cc}}$ with $t_{{\rm rh}}$ within $\approx10\%$ over
  three orders of magnitude in $N$. The scaling behavior becomes even better,
  with deviations $<1\%$ , if only the runs with $N\geq10^{6}$ are considered.
  The larger deviation in $t_{{\rm cc}}$ between the $N=10^{5}$ run and the
  other two are probably because of the larger stochastic variations in low $N$
  runs. 

\begin{figure} \includegraphics[scale=0.65]{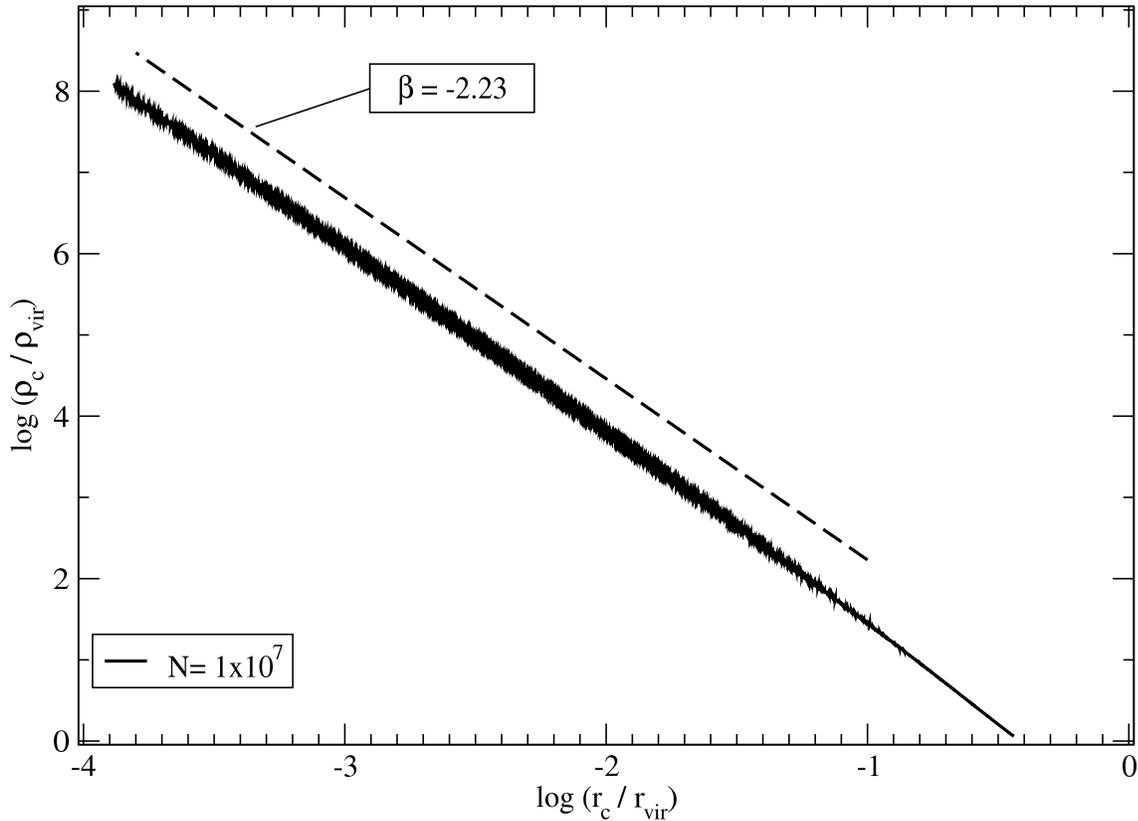}

  \caption{\label{fig:rhoc_rc_beta} Core density as function of $r_c/r_{vir}$.
    The density is expressed in units of $\rho_{vir}=M_c\,r^{-3}_{vir}$. The
    dashed line shows the power-law slope expected for a cluster close to core
    collapse based on the self-similarity of the cluster evolution in that
    regime \citep[see, e.g.,][]{2008gady.book.....B}. In our simulation this
    regime appears to be reached for $r_c/r_{vir}\lesssim 3\times 10^{-4}$, and
    a power-law fit to the core density in this range results in a slope of
  $-2.24$.}

\end{figure}

Another consequence of the self-similarity of the cluster evolution to core
collapse is that $-\beta\sim\log(\rho_{c}(t))/\log(r_{{\rm c}}(t))$
\citep{2008gady.book.....B}, which means that the decrease in $r_{{\rm c}}$
leads to an increase in $\rho_{{\rm c}}$ at a rate that is related to the shape
of the density profile. Indeed, from Figure \ref{fig:rcrh} one can see that the
shape of $\rho_{{\rm c}}(t)$ mirrors the shape of $r_{{\rm c}}(t)$ as expected.
Even more so, from Figure \ref{fig:rhoc_rc_beta}, we find that the
  power-law slope of $\rho_c(r_c)$ becomes $\beta=-2.24$ close to core-collapse
  ($r_c/r_{vir}<3\times 10^{-4}$), which is in excellent agreement with the
  corresponding $\beta=-2.23$ found by \citet{1980ApJ...242..765C}. It is worth
  noting that $\beta$ slowly changes with $r_c$, increasing from around
$\beta=-2.5$ and eventually converging to $\beta=-2.24$, which is also
reflected in the power-law slopes of the density profiles we obtain from our
simulations for increasing times (Figure \ref{fig:density_profile}).

Apart from the self-similar behavior, we also find that there is very little
mass loss ($\lesssim 1\%$), and hence very little energy is carried away by
escaping stars, in agreement with theoretical expectations
\citep[e.g.,][]{1978RvMP...50..437L}. Finally, we find that our code conserves
total energy to better than $0.04\%$ throughout the entire simulation.

\subsection{Performance Analysis\label{sub:Performance-Analysis}}

\begin{figure}
\includegraphics[scale=0.65]{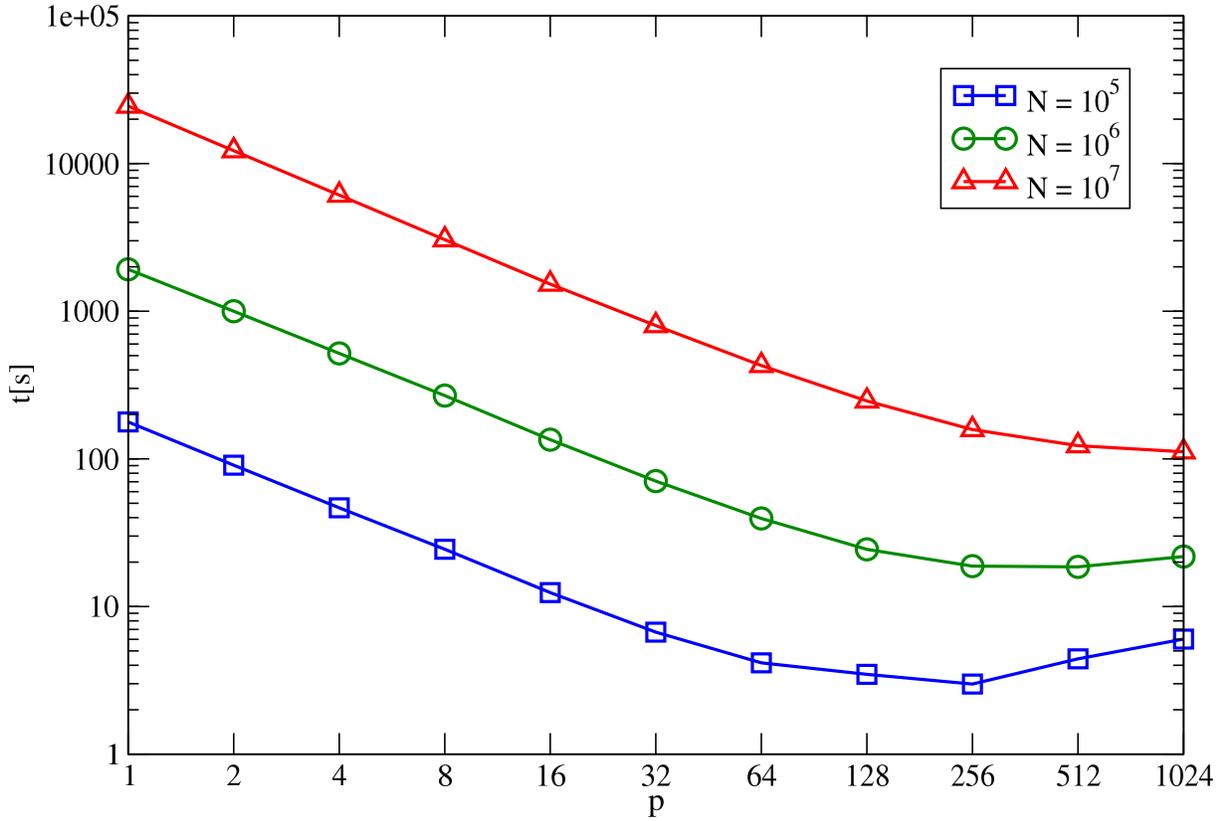}

\caption{\label{fig:Timing}Scaling of the wall-clock time with the number
of processors, $p$, for a parallel simulation of up to 50 timesteps. The
various lines represent different particle numbers (see legend). We observe a near-linear scaling for up to 64 processors.}
\end{figure}

We tested our parallel code for 3 cluster models with
$N$= $10^{5}$, $10^{6}$, and $10^{7}$ stars, with an initial
Plummer density profile. We used 1 to 1024 processors and measured
the total time taken, and time taken by various parts of the code
for up to 50 timesteps. For the sample size $s$ in the {\texttt Sample Sort} algorithm,
we chose 128, 256 and 1024 respectively, for the three $N$ values.

\begin{figure}
\includegraphics[scale=0.65]{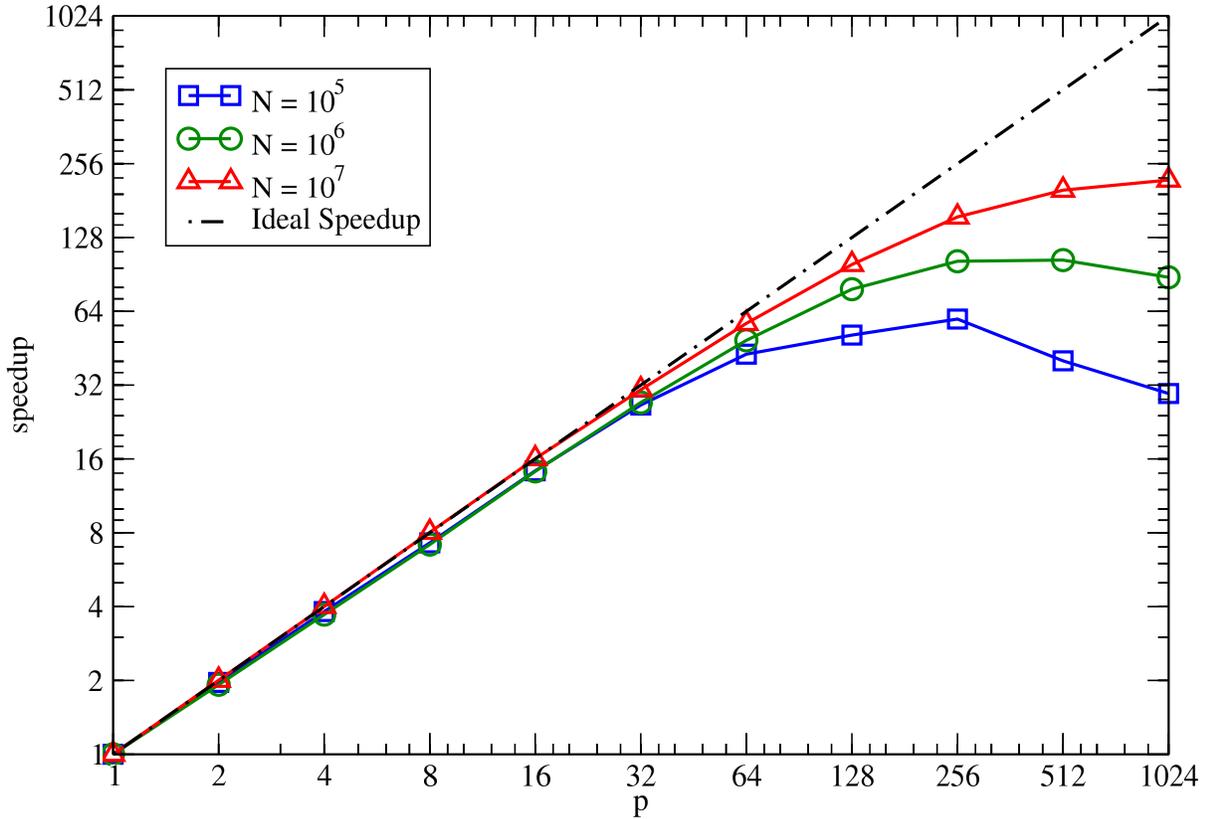}

\caption{\label{fig:Speedup}Speedup of our parallel code as a function of
the number of processors, $p$. The various lines represent different particle
numbers (see legend). The straight line represents the ideal speedup, which is the case when the speedup equals $p$. We observe that for all three cases, the speedup closely follows the ideal speedup line for up to 64 processors.}
\end{figure}

The timing results are shown in Figure \ref{fig:Timing} and a corresponding
plot of the speedups in Figure \ref{fig:Speedup}. These results do
not include the time taken for initialization. We can see that the speedup is nearly linear
up to 64 processors for all three runs, after which there is a gradual
decrease followed by saturation. For the $N=10^{5}$ and $10^{6}$
case, we also notice a dip after the saturation, also expected
for the $N=10^{7}$ case for a larger number of processors than we consider
here. We also see that the number of processors for which the speedup
peaks is different for each value of $N$, and gradually {\it increases} with $N$.
The peak is seen at 256 processors for the $N=10^{5}$ run, somewhere
between 256 and 512 for the $N=10^{6}$ run, and 1024 for the $N=10^{7}$
run. The maximum speedups observed are around $60\times$, $100\times$,
and $220\times$ for the three cases respectively. 

\begin{figure}
\includegraphics[scale=0.65]{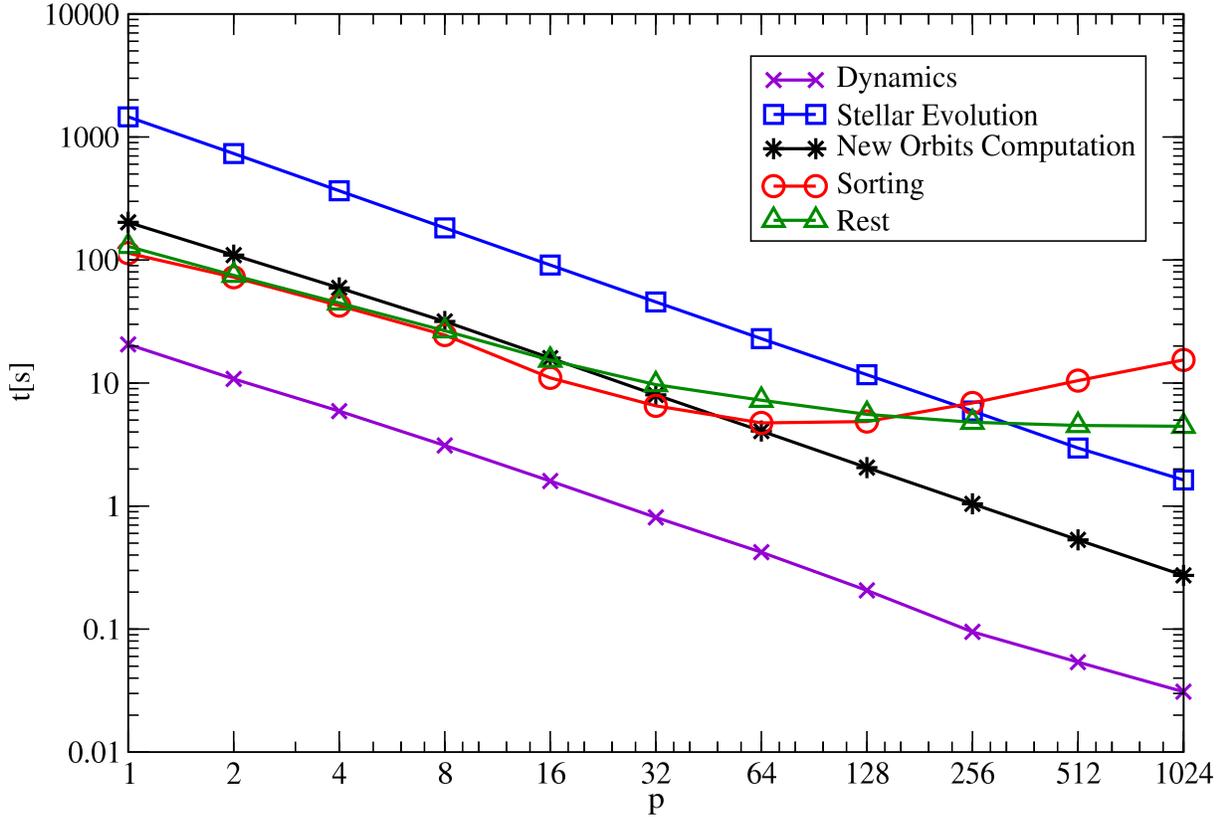}

\caption{\label{fig:modules_timing}Time taken by the various parts of the
code for a parallel run with $N=10^6$ stars. The various lines represent
the different modules of the algorithm (see legend). The module denoted by ``Rest" is the cumulative time taken by \emph{diagnostics}, \emph{potential calculation} and \emph{timestep calculation} steps. One can observe that the \emph{dynamics}, \emph{stellar evolution}, and \emph{new orbits computation} modules scale linearly whereas the other two exhibit relatively poor scaling beyond 64 processors.}
\end{figure}

Figure \ref{fig:modules_timing} shows the scaling of the time taken
by various modules of our code for the $N=10^6$ run. One can observe
that the dynamics, stellar evolution, and orbit calculation modules
achieve perfectly linear scaling. The ones that do not scale as well
are {\it sorting} and the ``rest", which include \emph{diagnostics}, \emph{potential calculation} and \emph{timestep calculation}. As the number
of processors increase, the linear scaling of the former three parts
of the code reduces their time to very small values, in turn letting
the parts that do not scale well dominate the runtime. This is the reason
for the trend observed in the total execution time and speedup plots.
We can also particularly see that the time taken by \emph{sorting} starts
to increase after reaching a minimum, and this explains a similar
observation in the previous plots as well.

\begin{figure}
\includegraphics[scale=0.65]{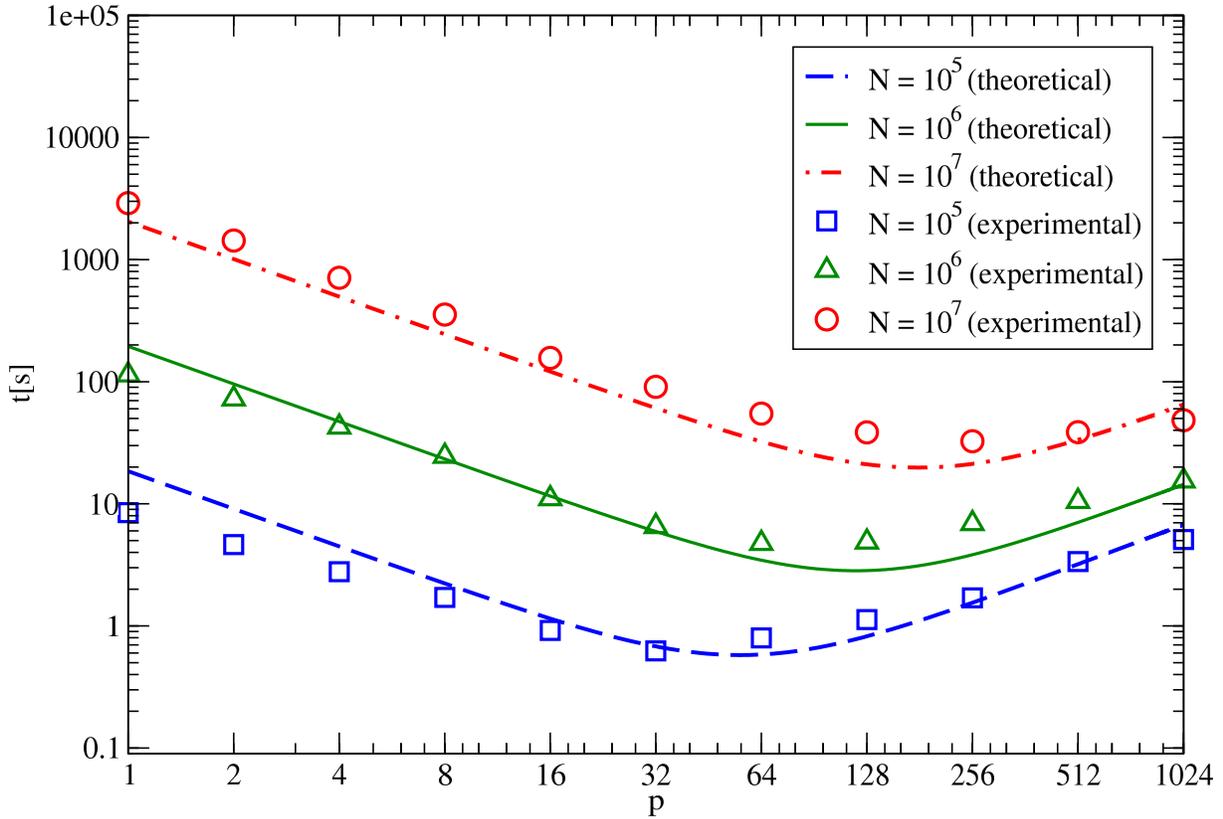}

\caption{\label{fig:model-sort}Time taken by the \emph{sorting} routine of the parallel
code plotted against the theoretical time complexity of {\texttt Sample Sort}
based on Equation \ref{eq:sort_complexity} for various values of
N. Appropriate proportionality constants were used based on the data size transported during the all-to-all communication phase of {\texttt Sample Sort} (see Section \ref{sub:Sorting}).}
\end{figure}

Figure \ref{fig:model-sort} shows the experimental timing results
of our \emph{sorting} step for the three $N$ values plotted against the
theoretical values calculated using Equation \ref{eq:sort_complexity}.
Since the entire star data is communicated during the all-to-all communication phase of the sort and not just the keys, and a data size of a single star is 46 times greater than that of
a single key in our code, we multiply the ${\cal O}(N/p)$ term of Equation
\ref{eq:sort_complexity} with a proportionality constant of $46$. We see that for all three cases,
the expected time linearly decreases, reaching a minimum after which
it increases. This implies that for every data size, there
is a threshold for the number of processors that would achieve optimal
performance beyond which it will worsen. The main reason is that, as the number of processors increases, smaller amounts of data are distributed across many processors, which makes the communication overhead dominant. In addition, the number of samples to be accumulated and sorted on a single node (Phase 2 and 3 of {\texttt Sample Sort}, see Section \ref{sub:Sorting}) increases with the number of processors, and hence this phase tends to become a bottleneck as well. From Figure \ref{fig:model-sort}, we also see that our implementation very closely follows the trend
predicted by the theoretical complexity formula. In addition, the
number of processors at which maximum performance is attained match
fairly closely as well. 

The other parts of the code that do not scale well include \emph{diagnostics}, \emph{potential calculation} and \emph{timestep calculation}. The computation of diagnostics and program termination related quantities require a number of collective communication calls to aggregate data across processors, and this is difficult to parallelize efficiently. However, there might be potential to improve this scaling by interleaving the collective communication calls with computation. While the \emph{timestep calculation} is embarrassingly parallel and can be expected to scale well, the time taken for potential calculation would remain constant irrespective of the number of processors used, since every processor computes the entire potential array, and has little scope for optimization. A thorough profiling of the poorly scaling parts is necessary in the future to identify the dominant among these bottlenecks, and prioritize optimization steps.

We noted earlier that the total execution time
 followed a very similar trend as \emph{sorting}, however, the places at which the minimum execution
time was observed are not the same, but shifted to the right. We can now infer this
to be a result of the linearly-scaling parts of the code pushing these points to the right until the relatively poorly scaling parts dominate the runtime. 

\section{Conclusions\label{sec:Conclusions}}

We presented a new parallel code, CMC (Cluster Monte Carlo), for simulating
collisional $N$-body systems with up to $N\sim10^{7}$. In order to maintain a
platform-independent implementation, we adopt the Message Passing Interface
(MPI) library for communication. The parallelization scheme uses a domain
decomposition that guarantees a near-equal distribution of data among
processors to provide a good balance of workload among processors, and at the
same time minimizes the communication requirements by various modules of the
code. Our code is based on the H\'enon Monte Carlo method, with algorithmic
modifications including a parallel random number generation scheme, and a
parallel sorting algorithm. We presented the first collisional $N$-body
simulations of star clusters with $N$ covering three orders of magnitude and
reaching up to $N=10^{7}$. The core collapse times obtained in our simulations
are in good agreement with previous studies, providing basic validation of our
code.  We also tested our implementation on 1 to 1024 processors. The code
scales linearly up to 64 processors for all cases considered, after which it
saturates, which we find to be characteristic of the parallel sorting
algorithm. The overall performance of the parallelization is impressive,
delivering maximum speedups of up to 220\texttimes{} for $N=10^{7}$.

Interesting future lines of work may include reducing the communication
overhead by overlapping communication with computation. In addition to the
distributed memory parallel version, CMC has also an optional feature that
accelerates parts of the algorithm using a general purpose Graphics Processing
Unit (GPU), described in the Appendix. An important next step towards reaching
even higher $N$ values is the development of a hybrid code which can run on
heterogeneous distributed architectures with GPUs. With progress along these
lines, we may be able to reach the domain of galactic nuclei for the first
time. Many galactic nuclei contain massive, dense star clusters, so-called
  nuclear star clusters, which are thought to be significantly linked to the
  evolution of their host galaxies \citep[e.g.,][]{2010IAUS..266...58B}, and
  their properties might still reflect to some extent the details of the
  formation of the galaxy \citep{2010ApJ...718..739M}. In addition, with their
  much larger masses and escape velocities, galactic nuclei are likely to
  retain many more stellar-mass black holes than globular clusters, and, thus,
  might significantly contribute to the black hole binary merger rate, as well
  as to the gravitational wave detection rate of advanced LIGO
  \citep{2009ApJ...692..917M}. Therefore, the study of galactic nuclei with a
  fully self-consistent dynamical code such as CMC has the potential to make
  strong predictions for future gravitational wave detection missions, and
might give further insights into the evolution of galaxies.

\acknowledgements{}

This work was supported by NASA Grant NNX09A036G to F.A.R. We also
acknowledge  partial support from NSF Grants PHY05-51164, CCF-0621443,
OCI-0724599, CCF-0833131, CNS-0830927, IIS- 0905205, OCI-0956311,
CCF-0938000, CCF-1043085, CCF-1029166, and OCI-1144061, and from DOE
Grants DE-FC02-07ER25808, DE-FG02-08ER25848, DE-SC0001283, DE-SC0005309,
and DE-SC0005340. This research used resources of the National Energy Research Scientific Computing Center, which is supported by the Office of Science of the U.S. Department of Energy under Contract No. DE-AC02-05CH11231.

\appendix{}

\section{GPU Acceleration of the Orbit Computation\label{sec:GPU-Implementation}}

An optional feature of the CMC code is the GPU acceleration of the
orbit computation that consists of finding $r_\text{min}$ and $r_\text{max}$ of each
stellar orbit and sampling a new orbital position (see Section \ref{sec:Code-Overview}).
As we have shown, the time complexity of both parts is ${\cal O}(N\log N)$,
and the orbit and new position for each star can be 
determined independently from the other stars. This makes the orbit computation
particularly well suited to being calculated on a GPU, not only because of
the inherent parallelism of the algorithm, but also for the large
number of memory accesses, which also scale as ${\cal O}(N\log N$), and,
thus, allow us to take advantage of the fast GPU memory.

Based on the structure of the algorithm, our implementation assigns
one thread on the GPU to do the computations for one star. This ensures
minimal data dependency between the threads since the same set of
operations are performed on different data, and makes the bisection
method and rejection technique implementations naturally suited for
SIMD (Single Instruction, Multiple Data) architectures, such as the
GPU. In the following we describe the specific adaptations of the
serial implementation of the algorithms to the GPU architecture and
present performance results.

\subsection{Memory Access Optimization}

To harness the high computation power of the GPU, it is very essential
to have a good understanding of its memory hierarchy in order to develop
strategies that reduce memory access latency. The first step towards
optimizing memory accesses is to ensure that memory transfer between
the host and the GPU is kept to a minimum. Another important factor
that needs to be considered is global memory coalescing in the GPU
which could cause a great difference in performance. When a GPU kernel
accesses global memory, all threads in groups of a half-warp access
a bank of memory at the same time \citep{NvidiaCorporation_2007}.
Coalescing of memory accesses happens when data requested by these
groups of threads are located in contiguous memory addresses, in which
case they can be read in one (or very few number of) access(es). Hence,
whether data is coalesced or not has a significant impact on an application\textquoteright{}s
performance as it determines the degree of memory access parallelism.
In CMC, the physical properties of each star are stored in a C structure,
containing 46 double precision variables. The N stars are stored in
an array of such C structures. 

\begin{figure} \includegraphics[width=1\columnwidth]{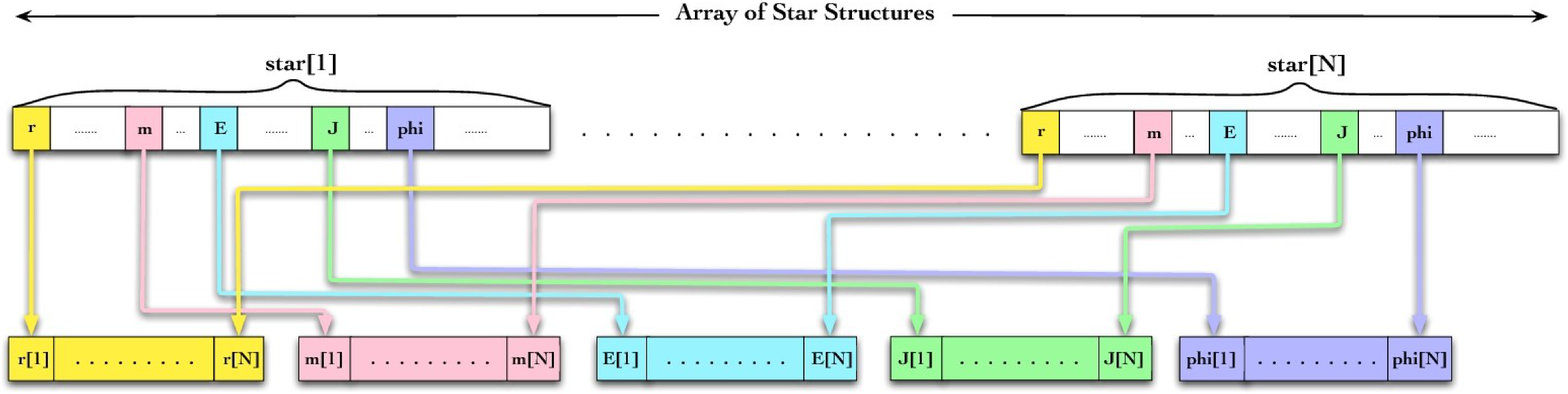}

\caption{\label{fig:Data-coalescing-strategy}Data coalescing strategy used
to strip the original star data structure and pack into contiguous
arrays before transferring them to the GPU. The original data layout is an array of structures (top), which is reorganized into fewer separated arrays (bottom) for the variables used in the new orbits computation.} 
\end{figure}
Figure \ref{fig:Data-coalescing-strategy} gives a schematic representation
of the data reorganization. At the top, the original data layout is
shown, i.e., an array of structures. The kernels we parallelize only
require 5 among the 46 variables present in the star structure: radial
distance, $r$, mass $m$, energy, $E$, angular momentum, $J$, and
potential at $r$, $\phi$, which are shown in different color. To
achieve coalesced memory accesses, we need to pack the data before
transferring it to the GPU in a way that they would be stored in contiguous
memory locations in the GPU global memory. A number of memory accesses
involve the same set of properties for different stars being accessed
together by these kernels since one thread works on the data of one
star. Hence, we extract and pack these into separate, contiguous arrays,
one for each property. This ensures that the memory accesses in the
GPU will be mostly coalesced. Also, by extracting and packing only
the 5 properties required by the parallel kernels, we minimize the
data transfer between the CPU and GPU.

\subsection{Random Number Generation}

For the generation of random numbers we use the same combined Tausworthe
generator and parallel implementation scheme as described in Section
\ref{sub:Parallel-Random-Number}. That is, for each thread that samples
the position of a star, there is one random stream with an initial
state that has been obtained by jumping multiple times in a seeded
random sequence to ensure statistical independence between streams.
As we will see later, to achieve optimal performance, 6000 to 8000
threads, and, thus, streams, are required. This can be easily accomodated
by one random sequence, as for our jump distance of $2^{80}$ and
a random number generator period of $\approx2^{113}$, $\approx10^{10}$
non-overlapping streams can be generated. Once generated on the host,
the initial stream states are transferred to the GPU global memory.
Each thread reads the respective starting state from the memory and
produces random numbers independently.

\subsection{Performance Results}

All our simulations are carried out on a 2.6 GHz AMD PhenomTM Quad-Core
Processor with 2 GB of RAM per core and an NVIDIA GTX280 GPU, with
30 multiprocessors, and 1 GB of RAM. The algorithms have been written
in the CUDA C language, and were compiled with the version 3.1 of
the CUDA compiler. All computations are done in double precision,
using the only Double Precision Unit (DPU) in each multiprocessor
on the GTX280 GPU.

We collect the timing results for 5 simulation timesteps of a single-mass
cluster with a Plummer density profile, and sizes ranging from $10^{6}$
to $7\times10^{6}$ stars, encompassing $\approx 25\%$ of all globular 
cluster sizes \citep[e.g.,][]{2005ApJS..161..304M}. 

\begin{figure}
\includegraphics[width=0.9\columnwidth]{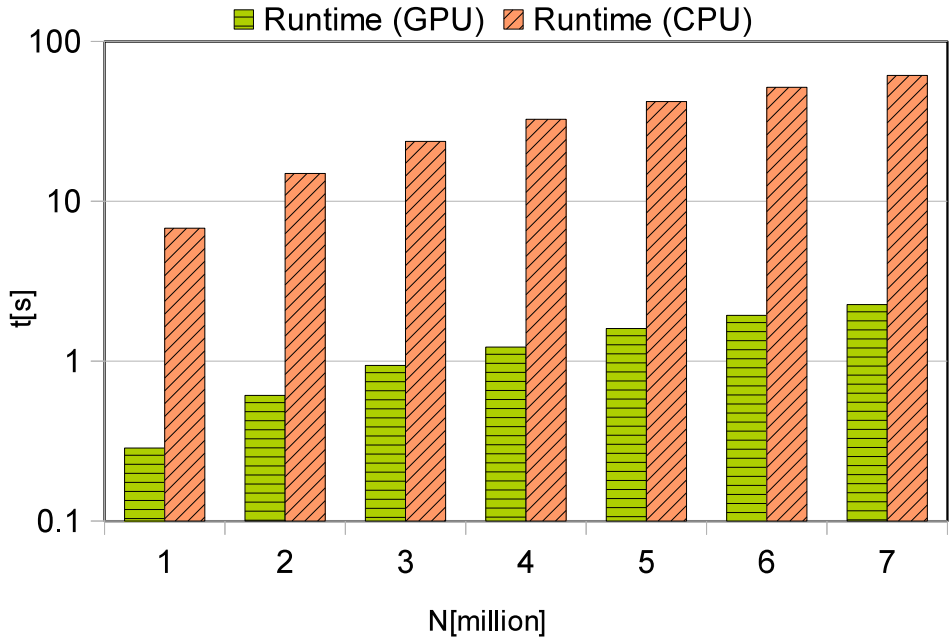}

\caption{\label{fig:Comparison-of-total-runtime}Comparison of total runtimes
of the sequential and parallelized kernels for various $N$.}
\end{figure}
Figure \ref{fig:Comparison-of-total-runtime} compares the GPU and
CPU runtimes. 
\begin{figure}
\includegraphics[width=0.9\columnwidth]{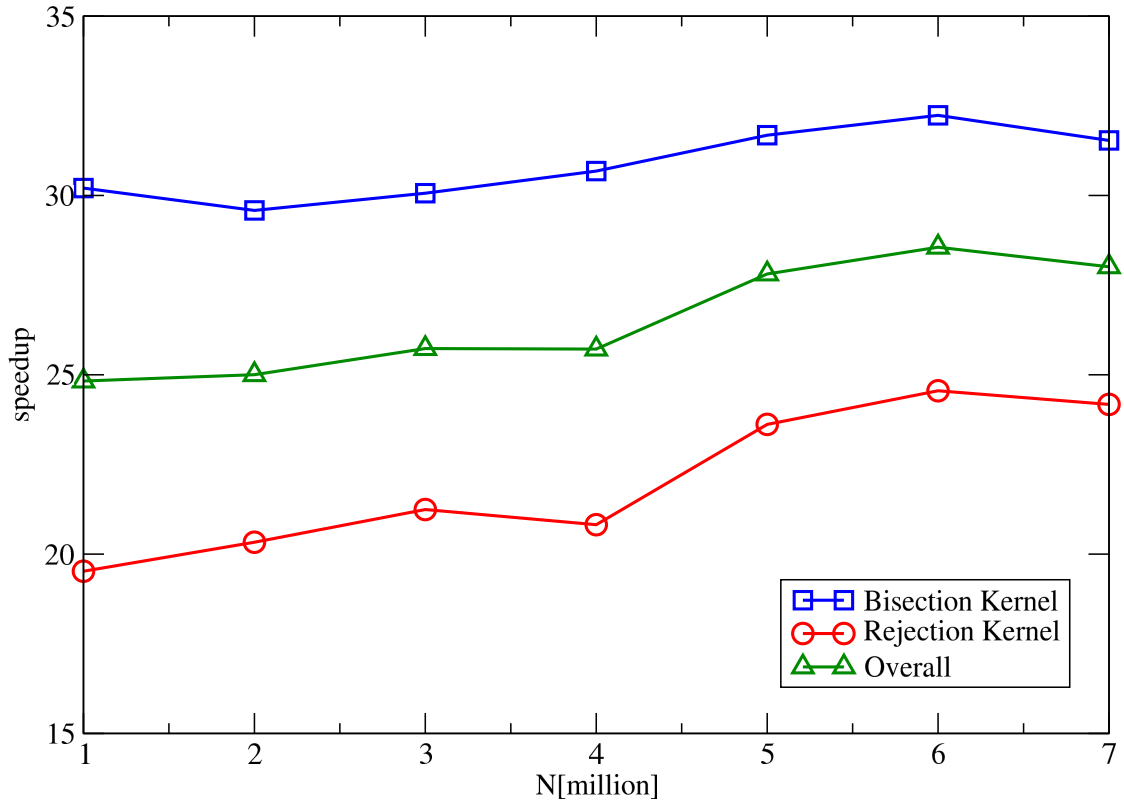}

\caption{\label{fig:Total-and-individual}Overall and individual speedups of
the bisection and rejection kernels. The mean overall speedup was observed to be $28\times$, with the individual mean speedups being $22\times$ and $31\times$ for the bisection and rejection kernels respectively. }
\end{figure}
Figure \ref{fig:Total-and-individual} shows the speedup of the \emph{new
orbits computation} part and the bisection and rejection
kernels individually. All speedup values are with respect to the code
performance on a single CPU. We see that the average speedups for
the rejection and bisection kernels are 22 and 31, respectively. This
is due to the difference in the number of floating point operations
between the two kernels which is a factor of 10. This makes a major
difference on the CPU but not on the GPU as it has more arithmetic
logic units (ALUs). Indeed, the bisection and rejection kernels take
about equal amount of time on the GPU for all $N$. This also indicates
that the performance of these kernels is only limited by the memory
bandwidth as they roughly require the same amount of global memory
accesses.

We also observe that the total speedup increases slightly as the data
size increases. 

In general, we obtain very good scalability. Analyzing the dependence
of the runtime on $N$ in Figure \ref{fig:Comparison-of-total-runtime}
we find that the GPU runtimes follow closely the kernel\textquoteright{}s
complexity of ${\cal O}(N\log N)$. The runtimes on the CPU, on the
other hand, have a steeper scaling with $N$, such that the run with
$N=7\times10^{6}$ takes a factor of $11$ longer than with $N=10^{6}$,
instead of the expected factor of 8. The reason for the somewhat worse
scaling of the runtime on the CPU is not yet clear and remains to
be investigated in the future. 

Note that as the memory transfer between the CPU and GPU is currently
not optimized, our speedup calculations do not include that overhead.
However, as we transfer only a subset of the entire data for each
star, there is the potential space for improvement to interleave kernel
computations with data transfer and substantially reduce this overhead. 

Finally, we looked at the influence of GPU specific parameters on
the runtime. In order to improve performance and utilization of the
$30$ multi-processors, we partitioned our data space into a one-dimensional
grid of blocks on the GPU. Due to the complexity of the expressions
involved in the calculations of orbital positions, our kernels use
a significant amount of registers (64 registers per thread). Thus,
the block dimension is restricted to 256 threads per block as the
GTX280 GPU has only 16384 registers per block. To analyze the performance,
we first made a parameter scan in the block and grid dimensions by
varying the block sizes from 64 to 256 and the grid sizes from 12
to 72. 
\begin{figure}
\includegraphics[width=0.9\columnwidth]{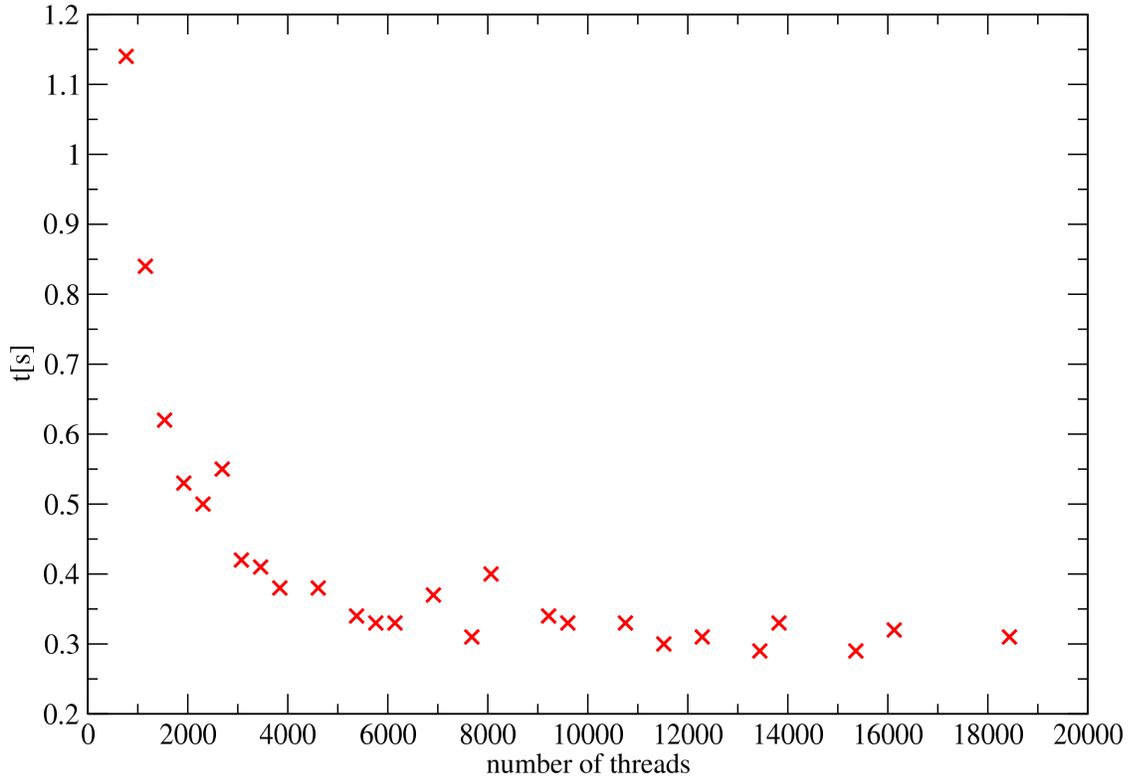}

\caption{\label{fig:Total-run-time-of}Total runtime of all kernels over the
total number of threads. The runtime decreases with increasing thread number and saturates at around 6000 threads, which is expected to be due to the finite memory bandwidth.}
\end{figure}
 Figure \ref{fig:Total-run-time-of} shows the total runtime of all
kernels as a function of the total number of active threads on the
GPU. As expected, the runtime decreases with increasing thread number
but saturates at around $6000$ threads. The saturation is most likely
due to the finite memory bandwidth, as we noted before that the runtime
of the kernels is mainly determined by the number of memory accesses
as opposed to the number of floating point operations. One can also
see that the curve shows little scatter, $\approx0.1\,{\rm s}$, which
means that the specific size of a single block of threads has a minor
effect on performance. We furthermore find that for a given total
number of threads, the runtime is shortest when the total number
of thread blocks is a multiple of the number of multi-processors,
in our case $30$.


\begin{thebibliography}{28}
\expandafter\ifx\csname natexlab\endcsname\relax\def\natexlab#1{#1}\fi

\bibitem[{{Binney} \& {Tremaine}(2008)}]{2008gady.book.....B}
{Binney}, J., \& {Tremaine}, S. 2008, {Galactic Dynamics: Second Edition}
  (Princeton University Press)

\bibitem[B{\"o}ker(2010)]{2010IAUS..266...58B} B{\"o}ker, T.\ 2010, IAU
Symposium, 266, 58 

\bibitem[Casertano 
\& Hut(1985)]{1985ApJ...298...80C} Casertano, S., \& Hut, P.\ 1985, \apj, 298, 80 

\bibitem[{{Chatterjee} {et~al.}(2010){Chatterjee}, {Fregeau}, {Umbreit}, \&
  {Rasio}}]{2010ApJ...719..915C}
{Chatterjee}, S., {Fregeau}, J.~M., {Umbreit}, S., \& {Rasio}, F.~A. 2010,
  \apj, 719, 915

\bibitem[{{Cohn}(1980)}]{1980ApJ...242..765C}
{Cohn}, H. 1980, \apj, 242, 765

\bibitem[{{Collins}(2008)}]{Army_report}
{Collins}, J.~C. 2008, Army Research Laboratory, 4, 41

\bibitem[{Fraser \& McKellar(1970)}]{samplesort}
{Fraser}, W.~D., {McKellar}, A.~C. 1970, JACM, 17, 496

\bibitem[{{Fregeau} {et~al.}(2004){Fregeau}, {Cheung}, {Portegies Zwart}, \&
  {Rasio}}]{2004MNRAS.352....1F}
{Fregeau}, J.~M., {Cheung}, P., {Portegies Zwart}, S.~F., \& {Rasio}, F.~A.
  2004, \mnras, 352, 1

\bibitem[{{Fregeau} {et~al.}(2003){Fregeau}, {G{\"u}rkan}, {Joshi}, \&
  {Rasio}}]{2003ApJ...593..772F}
{Fregeau}, J.~M., {G{\"u}rkan}, M.~A., {Joshi}, K.~J., \& {Rasio}, F.~A. 2003,
  \apj, 593, 772

\bibitem[{{Fregeau} \& {Rasio}(2007)}]{2007ApJ...658.1047F}
{Fregeau}, J.~M., \& {Rasio}, F.~A. 2007, \apj, 658, 1047

\bibitem[{{Freitag} \& {Benz}(2001)}]{2001A&A...375..711F}
{Freitag}, M., \& {Benz}, W. 2001, \aap, 375, 711

\bibitem[Giersz(1998)]{1998MNRAS.298.1239G} Giersz, M.\ 1998, \mnras, 298, 
1239 

\bibitem[Giersz et al.(2011)]{2011arXiv1112.6246G} Giersz, M., Heggie, D.~C.,
  Hurley, J., \& Hypki, A.\ 2011, arXiv:1112.6246 

\bibitem[{{Goswami} {et~al.}(2011){Goswami}, {Umbreit}, {Bierbaum}, \&
  {Rasio}}]{2011arXiv1105.5884G}
{Goswami}, S., {Umbreit}, S., {Bierbaum}, M., \& {Rasio}, F.~A. 2011, ArXiv
  e-prints


\bibitem[{{Gropp} {et~al.}(1994), {Lusk} \& {Skjellum}}]{mpibook}
{Gropp}, W., {Lusk}, E., \& {Skjellum}, A. 1994, {Using MPI:
    Portable Parallel Programming with the Message Passing Interface}
    (MIT Press)

\bibitem[Heggie \& Giersz(2009)]{2009MNRAS.397L..46H} Heggie, D.~C., \& Giersz,
  M.\ 2009, \mnras, 397, L46 

\bibitem[{{Heggie} \& {Hut}(2003)}]{2003gmbp.book.....H}
{Heggie}, D., \& {Hut}, P. 2003, {The Gravitational Million-Body Problem: A
  Multidisciplinary Approach to Star Cluster Dynamics}, ed. {Heggie, D.~\& Hut,
  P.}

\bibitem[{{Heggie} \& {Stevenson}(1988)}]{1988MNRAS.230..223H}
{Heggie}, D.~C., \& {Stevenson}, D. 1988, \mnras, 230, 223

\bibitem[{{H{\'e}non}(1971)}]{1971Ap&SS..14..151H}
{H{\'e}non}, M.~H. 1971, \apss, 14, 151

\bibitem[{Hoare(1961)}]{Hoare:1961:AQ:366622.366644}
Hoare, C. A.~R. 1961, Commun. ACM, 4, 321


\bibitem[{{Hurley} {et~al.}(2000){Hurley}, {Pols}, \&
  {Tout}}]{2000MNRAS.315..543H}
{Hurley}, J.~R., {Pols}, O.~R., \& {Tout}, C.~A. 2000, \mnras, 315, 543

\bibitem[{{Hurley} {et~al.}(2002){Hurley}, {Tout}, \&
  {Pols}}]{2002MNRAS.329..897H}
{Hurley}, J.~R., {Tout}, C.~A., \& {Pols}, O.~R. 2002, \mnras, 329, 897

\bibitem[Hurley \& Shara(2012)]{2012arXiv1208.4880H} Hurley, J.~R., \& Shara,
  M.~M.\ 2012, arXiv:1208.4880

\bibitem[{{Hut} {et~al.}(1988){Hut}, {Makino}, \&
  {McMillan}}]{1988Natur.336...31H}
{Hut}, P., {Makino}, J., \& {McMillan}, S. 1988, \nat, 336, 31

\bibitem[{{Jalali} {et~al.}(2012){Jalali}, {Baumgardt}, {Kissler-Patig},
  {Gebhardt}, {Noyola}, {L{\"u}tzgendorf}, \& {de Zeeuw}}]{2012A&A...538A..19J}
      {Jalali}, B., {Baumgardt}, H., {Kissler-Patig}, M., {et~al.} 2012, \aap,
      538, A19

\bibitem[{{Joshi} {et~al.}(2001){Joshi}, {Nave}, \&
  {Rasio}}]{2001ApJ...550..691J}
{Joshi}, K.~J., {Nave}, C.~P., \& {Rasio}, F.~A. 2001, \apj, 550, 691

\bibitem[{{Joshi} {et~al.}(2000){Joshi}, {Rasio}, \& {Portegies
  Zwart}}]{2000ApJ...540..969J}
{Joshi}, K.~J., {Rasio}, F.~A., \& {Portegies Zwart}, S. 2000, \apj, 540, 969

\bibitem[{L'Ecuyer(1999)}]{L'Ecuyer:1999:TME:307090.307103}
L'Ecuyer, P. 1999, Math. Comput., 68, 261

\bibitem[{Li {et~al.}(1993)Li, Lu, Schaeffer, Shillington, Wong, \&
  Shi}]{Li19931079}
Li, X., Lu, P., Schaeffer, J., {et~al.} 1993, Parallel Computing, 19, 1079

\bibitem[{{Lightman} \& {Shapiro}(1978)}]{1978RvMP...50..437L}
{Lightman}, A.~P., \& {Shapiro}, S.~L. 1978, Reviews of Modern Physics, 50, 437

\bibitem[{{Lusk} {et~al.}(1996){Doss}, \& {Skjellum}}]{mpibasic}
{Lusk}, E., {Doss}, N., \& {Skjellum}, A. 1996, Parallel Computing, 22, 789



\bibitem[{{McLaughlin} \& {van der Marel}(2005)}]{2005ApJS..161..304M}
{McLaughlin}, D.~E., \& {van der Marel}, R.~P. 2005, \apjs, 161, 304

\bibitem[Merritt(2010)]{2010ApJ...718..739M} Merritt, D.\ 2010, \apj, 718, 
739

\bibitem[{{Miller} \& {Lauburg}(2009)}]{2009ApJ...692..917M}
{Miller}, M.~C., \& {Lauburg}, V.~M. 2009, \apj, 692, 917

\bibitem[{Nvidia(2010)}]{NvidiaCorporation_2007}
Nvidia. 2010, {http://developer.download.nvidia.com}, 1

\bibitem[Spitzer(1987)]{1987degc.book.....S} Spitzer, L.\ 1987, Princeton, NJ,
  Princeton University Press, 1987, 191 p.,  

\bibitem[Stodolkiewicz(1982)]{1982AcA....32...63S} Stodolkiewicz, J.~S.\ 
1982, \actaa, 32, 63 

\bibitem[{{Takahashi}(1995)}]{1995PASJ...47..561T}
{Takahashi}, K. 1995, \pasj, 47, 561

\bibitem[Trenti et al.(2010)]{2010ApJ...708.1598T} Trenti, M., Vesperini, E.,
  \& Pasquato, M.\ 2010, \apj, 708, 1598 

\bibitem[{{Umbreit} {et~al.}(2012){Umbreit}, {Fregeau}, {Chatterjee}, \&
  {Rasio}}]{2012ApJ...750...31U}
{Umbreit}, S., {Fregeau}, J.~M., {Chatterjee}, S., \& {Rasio}, F.~A. 2012,
  \apj, 750, 31

\bibitem[{{Zonoozi} {et~al.}(2011){Zonoozi}, {K{\"u}pper}, {Baumgardt},
  {Haghi}, {Kroupa}, \& {Hilker}}]{2011MNRAS.411.1989Z}
{Zonoozi}, A.~H., {K{\"u}pper}, A.~H.~W., {Baumgardt}, H., {et~al.} 2011,
  \mnras, 411, 1989

\end{thebibliography}
\end{document}